\documentclass[11pt]{article}

\usepackage{graphicx}
\usepackage{amssymb}
\usepackage{amsmath}
\usepackage{enumerate}
\usepackage{amsfonts}
\usepackage[section]{placeins}
\usepackage[colorlinks=true,linktocpage=true,linkcolor=blue,citecolor=blue]{hyperref}

\newcommand{\be}{\begin{equation}}
\newcommand{\ee}{\end{equation}}
\newcommand{\bea}{\begin{eqnarray}}
\newcommand{\eea}{\end{eqnarray}}
\newcommand{\vev}[1]{{\left< {#1} \right>}}

\newcommand{\eqn}[1]{(\ref{#1})}

\newcommand{\mt}[1]{\textrm{\tiny #1}}
\def\nc {N_\mt{c}}
\def\nf {N_\mt{f}}

\newcommand{\pa}{\partial}

\newcommand{\sac}{\, , \qquad}

\newcommand{\uh}{u_\mt{H}}
\newcommand{\phib}{\phi_\mt{bdry}}
\newcommand{\phit}{\tilde{\phi}}
\newcommand{\phith}{\tilde{\phi}_\mt{H}}
\newcommand{\phih}{{\phi}_\mt{H}}
\newcommand{\Bb}{{\cal B}_\mt{bdry}}
\newcommand{\Fb}{{\cal F}_\mt{bdry}}
\newcommand{\Hb}{{\cal H}_\mt{bdry}}
\newcommand{\Bh}{{\cal B}_\mt{H}}
\newcommand{\g}{\gamma}
\newcommand{\cs}{c_\mt{sch}}
\newcommand{\ra}{\rightarrow}

\def\pa{\partial}                                       
\newcommand{\gym}{g_\mt{YM}}
\newcommand{\mq}{M_\mt{q}}
\newcommand{\cc}{{\cal C}}
\newcommand{\cf}{{\cal F}}
\newcommand{\cb}{{\cal B}}
\newcommand{\ch}{{\cal H}}
\newcommand{\cn}{{\cal N}}
\newcommand{\n}{n_\mt{D7}}

\begin{document}
\bibliographystyle{hieeetr}

\pagestyle{plain}
\setcounter{page}{1}

\begin{titlepage}

\begin{center}
\vspace*{-1cm} \today \hfill ICCUB-11-145  \\
{}~{} \hfill  MAD-TH-11-03 \\

\vskip 3cm

{\LARGE {\bf Thermodynamics and Instabilities}}

\vskip 3mm

{\LARGE {\bf  of a}}

\vskip 3.5mm

{\LARGE {\bf  Strongly Coupled Anisotropic Plasma}}

\vskip 1 cm

{\large {\bf David Mateos$^{1,2}$ and Diego Trancanelli$^{3,4}$}}

\vspace{1.5mm}

\vskip .8cm
  ${}^1$ {\it Instituci\'o Catalana de Recerca i Estudis Avan\c cats (ICREA), Passeig Llu\'\i s Companys 23, E-08010, Barcelona, Spain}

\medskip

  ${}^2$ {\it Departament de F\'\i sica Fonamental (FFN) \&  Institut de Ci\`encies del Cosmos (ICC), Universitat de Barcelona (UB), Mart\'{\i}  i Franqu\`es 1, E-08028 Barcelona, Spain}

\medskip

  ${}^3$ {\it Department of Physics, University of Wisconsin, Madison, WI 53706, USA}

\medskip 

  ${}^4$ {\it Department of Physics, University of California, Santa Barbara, CA 93106, USA}

\vskip 0.5cm

{\tt  dmateos@icrea.cat, \, dtrancan@physics.wisc.edu}

\vspace{5mm}

{\bf Abstract}
\end{center}
 \noindent
We extend our analysis of a IIB supergravity solution dual to a spatially anisotropic finite-temperature ${\cal N}=4$ super Yang-Mills plasma. The solution is static, possesses an anisotropic horizon, and is completely regular. The full geometry can be viewed as a renormalization group flow from an AdS geometry in the ultraviolet to a Lifshitz-like geometry in the infrared. The anisotropy can be equivalently understood as resulting from a position-dependent $\theta$-term or from a non-zero number density of dissolved D7-branes. The holographic stress tensor is conserved and anisotropic. The presence of a conformal anomaly plays an important role in the thermodynamics. The phase diagram exhibits homogeneous and inhomogeneous (i.e.~mixed) phases. In some regions the homogeneous phase displays instabilities reminiscent of those of weakly coupled plasmas.  We comment on similarities with QCD at finite baryon density and with the phenomenon of cavitation.
\end{titlepage}

\tableofcontents


\section{Introduction}
A remarkable conclusion from the Relativistic Heavy Ion Collider (RHIC) experiments \cite{rhic,rhic2} is that the quark-gluon plasma (QGP) does not behave as a weakly coupled gas of quarks and gluons, but rather as a strongly coupled fluid \cite{fluid,fluid2}. This renders perturbative methods inapplicable in general. The lattice formulation of Quantum Chromodynamics (QCD) is also of limited utility, since for example it is not well suited for studying real-time phenomena. This has provided a strong motivation for understanding the dynamics of strongly coupled non-Abelian plasmas through the gauge/string duality \cite{duality,duality2,duality3} (see \cite{review} for a recent review of applications to the QGP). The simplest example of the duality is the equivalence between four-dimensional ${\cal N}=4$ super Yang-Mills (SYM) theory with gauge group $SU(\nc)$ and type IIB string theory on $AdS_5 \times S^5$. 
We recently extended this duality to the case in which the SYM plasma is \emph{spatially} anisotropic \cite{prl}.\footnote{The term `anisotropic' is sometimes used in the literature to refer to systems in which there is an asymmetry between time and space, i.e.~systems with no Lorentz invariance. In this paper we will use the term `anisotropic' to refer exclusively to spatially anisotropic systems, i.e.~to systems in which there is an asymmetry between different spatial directions.} In this paper we provide additional details and extend our analysis of the anisotropic supergravity solution. Previous holographic studies of anisotropic plasmas include solutions dual to non-commutative SYM theories at finite temperature \cite{nc2,nc3,nc4,nc5,nc6}, gravity duals of p-wave superconductors\cite{pwave,pwave2,pwave3,pwave4,pwave5,pwave5bis,pwave6}, solutions dual to ${\cal N}=4$ SYM theory coupled to electromagnetic fields with \cite{with,with2,with3,with4} or without  \cite{B,B2} flavour, and anisotropic solutions with singular horizons \cite{sing}. 

Part of the motivation for our work comes from the fact that the plasma created in a heavy ion collision is anisotropic. The system is locally anisotropic for some short time after the collision, $\tau < \tau_\mt{iso}$, and becomes locally isotropic afterwards. An intrinsically anisotropic hydrodynamical description has been proposed to describe the early stage after the collision \cite{ani,ani2,ani3,ani4,ani5,ani6,ani7,ani8,ani9}, in which the plasma is assumed to have unequal pressures in the longitudinal and transverse directions. We will encounter this feature too. At a time $\tau_\mt{iso}$ the pressures become approximately equal and a standard hydrodynamic description can be applied in which the stress tensor is assumed to be locally isotropic, meaning that each little cube of QGP is isotropic in its own rest frame. Even in this phase, certain observables may be sensitive to the physics in several adjacent cubes, in which case there is no frame in which they can be computed as if no anisotropy were present. A simple example is the drag experienced by a heavy quark moving through the plasma. On the gravity side the quark is modeled by an infinitely long string \cite{drag,drag2}, which is therefore sensitive to the physics in a large region of the plasma. On the gauge theory side the size of the string corresponds roughly to the size of the gluon cloud that dresses the quark. 

Further motivation for our work is provided by the fact that weakly coupled  plasmas, both Abelian and non-Abelian, are known to generically suffer from instabilities in the presence of anisotropies \cite{Weibel:1959zz,Mrowczynski:1988dz,Mrowczynski:1993qm,Mrowczynski:1994xv,Mrowczynski:1996vh}. In fact, it has been proposed  \cite{Randrup:2003cw,Romatschke:2003ms,Arnold:2003rq,Romatschke:2004jh,Arnold:2004ti,Rebhan:2004ur,Arnold:2005vb} that these instabilities might be responsible for the seemingly small value of $\tau_\mt{iso} \lesssim 1$ fm. It is therefore interesting to understand whether this type of instabilities also occur in anisotropic plasmas at strong coupling. We will see that the gravity solution that we will present  exhibits instabilities reminiscent of weak-coupling instabilities. 

Another interesting aspect of our solution also related to the QGP is the fact that it is sourced by a position-dependent $\theta$-term --- see eqn.~\eqn{def} below. It has been suggested that `bubbles' or regions with a spacetime-dependent theta-angle, $\theta(t,x)$, may be formed in heavy ion collisions \cite{effective,effective2,effective3,effective4,cme,cme2}. In QCD an effective $\theta$-angle of this form  is equivalent to a non-zero axial vector potential $A_\mu^5=\partial_\mu \theta/2\nf$ that couples to the axial current as $\bar \psi \gamma^\mu \gamma^5 \psi A_\mu^5$. This leads to interesting effects, including the generation of an electric current along a magnetic field --- the so-called chiral magnetic effect \cite{effective,effective2,cme,cme2}.

We emphasize that the purpose of this paper is to study in detail the anisotropic SYM plasma at strong coupling through its dual gravity description. Although we will comment on superficial similarities and differences with the physics of the QGP, an in-depth analysis of the extent (if any) to which a close connection with the QGP exists is beyond the scope of this paper. For example, we leave for future work questions such as whether the instabilities we will identify can be thought of as a strong-coupling counterpart of the weak-coupling instabilities mentioned above.

At a more theoretical level, motivation for our work is provided by the fluid/gravity correspondence \cite{fluidgravity, fluidgravity1,fluidgravity2, fluidgravity3, fluidgravity4, fluidgravity5, fluidgravity6, fluidgravity7} (see \cite{rev,rev2,rev3} for reviews), which asserts that the long-wavelength effective dynamics of black brane horizons in asymptotically anti-de Sitter spacetimes is equivalent  to the long-wavelength effective dynamics of the dual finite-temperature plasma, namely to fluid dynamics. It would certainly be interesting to extend this correspondence to the case in which the fluid is allowed to be intrinsically anisotropic, for which purpose our solution may provide a starting point. 

An interesting connection also exists between our work and the so-called blackfold approach \cite{blackfold,blackfold2,blackfold3,blackfold4} to black hole dynamics, of which the fluid/gravity correspondence may be viewed as a holographic implementation. In this framework the long-wavelength dynamics of a black hole horizon is described by a hydrodynamic theory living on a dynamical worldvolume. The black hole is embedded in an arbitrary spacetime, which in particular is not necessarily asymptotically AdS, and may carry dipole charges which can be thought of as associated with $p$-dimensional objects dissolved in the black hole's horizon. In these cases the corresponding effective dynamics describes an anisotropic fluid with a conserved $p$-brane charge \cite{blackfold4}. As we will see, the anisotropy in the SYM plasma that we will study can be thought of as resulting from a dissolved 2-brane charge. 

A final motivation for our work comes from the possible applications to condensed matter systems.  In particular, our solution may be viewed as an ultraviolet (UV) completion of a charged, anisotropic Lifshitz-like infrared (IR) geometry, as already observed by the authors of Ref.~\cite{ALT}. This allows for a rigorous holographic study of the Lifshitz geometry and provides a microscopic description of the dual theory.\footnote{Embeddings of Lifshitz geometries in string theory were constructed in \cite{embed,embed2,embed3,embed4,embed5,linear,embed6,embed7,embed8}. Recent studies of AdS completions of charged, spatially isotropic Lifshitz geometries include \cite{comp,comp2, comp2bis,comp3,comp4,comp5,comp6, comp7}.} More generally, anisotropic strongly coupled condensed matter systems exist in Nature (e.g.~liquid crystals) and we hope that our work may contribute to the application of holographic methods to this type of systems.  

The type IIB supergravity solution that we will present is a finite-temperature generalization of that of Ref.~\cite{ALT}, which provided much inspiration for our work. In the Einstein frame the ten-dimensional geometry factorizes as ${\cal M} \times {\cal X}$, where ${\cal X}$ is a five-dimensional Einstein manifold which we will take to be ${\cal X}=S^5$ for simplicity, and ${\cal M}$ satisfies the following properties: 
\begin{enumerate}
\item
It is static and anisotropic (along the gauge theory directions).
\item
It possesses a horizon and it is regular on and outside the horizon.
\item
It approaches AdS$_5$ asymptotically.
\end{enumerate}
The staticity requirement is motivated by the desire to work in the simplest possible setup. In other words, we wish to be able to study the thermodynamics of the system, its response when it is slightly perturbed, etc. The presence of a horizon is dual to the existence of a finite-temperature plasma in the gauge theory. The requirement of regularity guarantees that calculations are unambiguous and well defined. Finally, the AdS boundary conditions ensure that holography is on its firmest footing. More specifically, the fact that our configuration solves the type IIB supergravity equations of motion and asymptotically approaches $AdS_5 \times S^5$ implies that it is dual to ${\cal N}=4$ SYM theory deformed by a (marginally) relevant operator, and thus that it is solidly embedded in string theory. 

As in \cite{ALT}, the deformation of the ${\cal N}=4$ theory that we consider corresponds to the addition of a $\theta$-parameter that depends linearly on one of the three spatial coordinates, $\theta = 2\pi \n z$, where  
$\{t,x,y,z\}$ are the gauge theory coordinates and $\n$ is a constant with dimensions of energy which (as we will review) can be interpreted as a density of D7-branes distributed along the $z$-direction. In other words, the total gauge theory action takes the form
\be
S_\mt{gauge}=S_{{\cal N}=4} + \delta S \sac 
\delta S = \frac{1}{8\pi^2} \int   \theta(z) \, \mbox{Tr} \, F \wedge F \,.
\label{def}
\ee
This clearly breaks isotropy (and CP) but not translation invariance, since integration by parts yields 
\be
\delta S \propto - \n \int  dz \wedge \mbox{Tr} \left( A \wedge F + \frac{2}{3} A^3 \right) \,. 
\ee
Incidentally, this expression shows that, if the $z$-direction is compactified on a circle, the resulting three-dimensional theory contains a Chern-Simons term. If in addition antiperiodic boundary conditions are imposed for the fermions around the circle, then the theory flows at long distances to a Chern-Simons theory \cite{QHE}. The deformation \eqn{def} breaks all the supersymmetries of the four-dimensional SYM theory. Supersymmetry-preserving deformations with a space-dependent  $\theta$-angle (and coupling constant) have been considered in \cite{Gaiotto:2008sd}, motivated by the construction of  Chern-Simons theories with ${\cal N}=4$ supersymmetry in three dimensions.

\begin{figure}
\begin{center}
\includegraphics[scale=0.8]{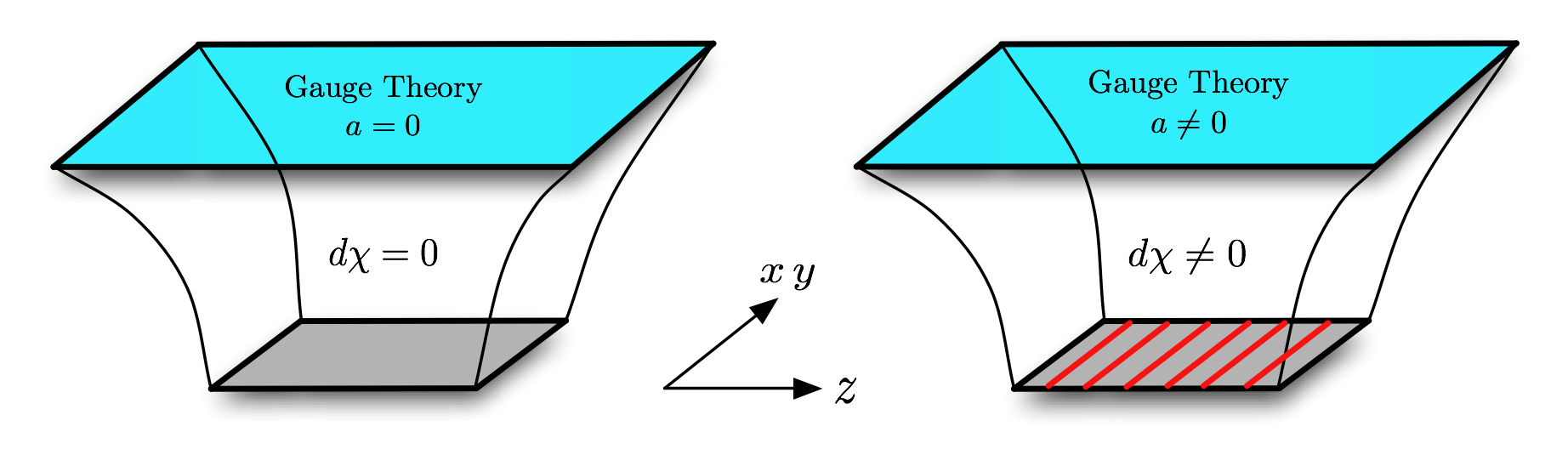}
\caption{\small D7-branes dissolved in the geometry.
\label{dissolved}}
\end{center}
\end{figure}

The complexified coupling constant of the ${\cal N}=4$ theory is related to the axion-dilaton of type IIB supergravity through
\be
\tau = \frac{\theta}{2\pi} + \frac{4\pi i}{\gym^2} = \chi + i e^{-\phi}\,.
\ee
Thus we expect that the gravity solution dual to the deformation \eqn{def} will have a position-dependent axion field of the form $\chi =a z$ where, as we will see, the constant $a$ is given by
\be
a=\frac{\lambda \n}{4\pi \nc} \,,
\label{a0}
\ee
where $\lambda=\gym^2 \nc$ is the gauge theory 't Hooft coupling.\footnote{IIB supergravity solutions with a running axion-dilaton were constructed in \cite{run,run2,run3}. Lifshitz-like solutions with a linear axion have  been considered in \cite{linear,embed7,embed8}.} Since the axion is magnetically sourced by D7-branes, this suggests that it should be possible to interpret the solution in terms of a number of D7-branes dissolved in the geometry \cite{ALT} --- see 
Fig.~\ref{dissolved}. 
We will elaborate on this in Section \ref{sec-thermodynamics}.  Here we just anticipate that the D7-branes are wrapped on the $S^5$ factor of the geometry, extend along the $xy$-directions and are homogeneously distributed along the $z$-direction with uniform density $n_\mt{D7} = d N_\mt{D7}/dz$. In view of the relation \eqn{a0} we will use $\n$ and $a$ interchangeably to refer to the D7-brane number or charge density. The orientation of the D3- and D7-branes that give rise to our solution can be summarized in a standard notation by the array
\bea
\begin{array}{c| cccc|c|c}
& t & x & y & z & u & S^5 \\
\hline 
\nc ~ \mbox{  \,  D3 } & \times & \times & \times & \times & & \\
 n_\mt{D7} ~ \mbox{  D7 } &  \times &  \times & \times & & &  \times 
\end{array}\,,
\eea
where $u$ is the holographic radial coordinate in AdS$_5$. 
We stress that, since our solution incorporates their full back-reaction, the D7-branes are completely `dissolved' in the geometry, just like the $\nc$ D3-branes that give rise to $AdS_5 \times S^5$. We also emphasize that, unlike the case of  D7-branes used to introduce flavour (quark) degrees of freedom in  
${\cal N}=4$ SYM \cite{flavour,flavour2,flavour3}, the D7-branes considered here do not extend in the radial direction. Consequently, they do not reach the AdS boundary and they do not add new degrees of freedom to the SYM theory.

We will verify that translation invariance is indeed preserved by showing that the holographic stress tensor is conserved. The calculation of the stress tensor will require that we  renormalize the gravity action by adding appropriate counterterms, which have been determined in \cite{odin,Yiannis}. In the process of renormalization we will discover that the deformation \eqn{def} induces a non-zero conformal anomaly, $\langle T^i_i \rangle \neq 0$. 
This will play an important role in the thermodynamics of the system, most crucially because it will imply the existence of an arbitrary reference scale $\mu$, a remnant of the renormalization process much like the subtraction point in QCD. This in turn means that \emph{the physics does not only depend on $a/T$}, but on two independent dimensionless ratios which may be taken to be $a/\mu$ and $T/\mu$. A detailed study of the thermodynamics of asymptotically locally AdS spacetimes, in particular in the presence of a holographic conformal anomaly \cite{anomaly,anomaly2}, was presented in \cite{thermo}.

As mentioned above, in the UV our solution approaches $AdS_5 \times S^5$, as appropriate for a marginally relevant deformation of the $\cn =4$ theory. In this limit the solution is of course invariant under the rescaling $x^i \ra k x^i, u\ra k u$. We will see that in the IR it approaches a Lifshitz-like solution whose metric is invariant under a scaling as above for all the coordinates except for $z$, which scales as $z \ra k^{2/3} z$. In this sense the solution may be viewed as a renormalization group (RG) flow between an isotropic UV fixed point and an anisotropic IR fixed point.\footnote{This is a slight abuse of language because, although the metric in the IR is scale-invariant, the dilaton and the Ramond-Ramond forms are not. See Appendix \ref{comparison}.} The zero-temperature version of this flow was found in \cite{ALT}, and in this case the string-frame metric exhibits a naked curvature singularity deep in the IR (see Section \ref{discussion} and Appendix \ref{comparison} for details). However, we emphasize that at finite temperature the singularity is hidden behind the horizon and the solution is completely regular on and outside the horizon, exhibiting no pathologies regardless of the number of D7-branes, which in particular can be larger than 24. 
Incidentally, this provides a completely explicit counterexample (at least in the presence of a negative cosmological constant) to the somewhat extended belief that 24 is an upper bound on the possible number of D7-branes. We will comment further on this point below. The finite-temperature deformation of the IR Lifshitz geometry was also found analytically by the authors of \cite{ALT}, which allowed them to show that in this case the entropy density scales as $s\sim a^{1/3} T^{8/3}$. 
Holographic renormalization for supergravity solutions dual to field theory RG flows was initiated in \cite{kostas,kostas2}.

The organization of the paper is as follows. In Section~\ref{action-section} we review the solution presented in \cite{prl} and discuss the entropy density. In Section~\ref{stress-section} we compute the stress tensor particularizing the results of \cite{odin,Yiannis}, and discuss the role of the conformal anomaly. In Section~\ref{sec-thermodynamics} we discuss the definition of the chemical potential conjugate to the D7-brane number density $\n$ and review several aspects of the thermodynamics of anisotropic systems. In Section~\ref{sec-phasediagram} we apply the results from previous sections to the construction of the phase diagram in the $(a,T)$-plane (i.e.~in the canonical ensemble). We identify regions in which the homogeneous anisotropic phase is stable, metastable or unstable. In the latter two cases the stable configuration is a mixed (inhomogeneous) configuration in which the anisotropic and the isotropic phases coexist. A summary of our results is contained in Section~\ref{discussion}, where we also elaborate on several topics including the similarities and differences of our system with weakly coupled plasmas and with QCD at finite baryon density.

We conclude the paper with a series of appendices in which we collect some of the more technical details of our computations. In Appendix~\ref{derivation} we write down the equations of motion and describe the procedure we have used to solve them. In Appendix~\ref{comparison} we compare our parametrization of the geometry with the parametrization employed in \cite{ALT}. In Appendix~\ref{FPrelations} we derive the relations between thermodynamic potentials and pressures in an anisotropic system. 
In Appendices~\ref{highApp} and \ref{lowApp} we obtain analytic expressions for the metric and the various thermodynamical quantities in the regimes $T\gg a$ and $T\ll a$, respectively.


\section{Action and solution}
\label{action-section}
In this section we collect the main results about the solution of interest. Additional details can be found in Appendix \ref{derivation}.

The ten-dimensional solution we seek is a direct product of the form ${\cal M} \times S^5$, where the radius $L$ of the  $S^5$ is constant in the Einstein frame and is given by
\be
L^4 = 4\pi g_s \nc \ell_s^4 = \lambda \ell_s^4 \,,
\ee
where $\lambda = \gym^2 \nc$ is the gauge theory `t Hooft coupling. Therefore ${\cal M}$  can be viewed as a solution of five-dimensional supergravity with a negative cosmological constant $\Lambda = -6/L^2$. Since only the metric $g_{\mu\nu}$, the axion $\chi$, and the dilaton $\phi$ will be excited, it suffices to consider the five-dimensional axion-dilaton-gravity action, which in the Einstein frame takes the form
\be
S_\mt{bulk} = \frac{1}{2 \kappa^2} \int_{\cal M} \sqrt{-g} \left( R + 12 -\frac{1}{2} (\partial \phi)^2 
-\frac{1}{2} e^{2\phi} (\partial \chi)^2 \right) 
+\frac{1}{2 \kappa^2} \int_{\cal \partial M} \sqrt{-\gamma} \, 2 K \,,
\label{action}
\ee
where $2\kappa^2 = 16 \pi G$ is the five-dimensional gravitational coupling. For convenience we will measure all lengths in units of $L$, i.e.~we have set $L=1$ in eqn.~\eqn{action}, which implies that $G=\pi/2\nc^2$ and $\kappa^2=4\pi^2/\nc^2$ (see e.g.~\cite{review}). The  solution takes the form
\bea
&&\hskip -.35cm 
ds^2 =  \frac{e^{-\frac{1}{2}\phi}}{u^2}
\left( -\cf \cb\, dt^2+dx^2+dy^2+ \ch dz^2 +\frac{ du^2}{\cf}\right), 
\,\,\,\,\,\,\label{sol1} \\
&& \hskip -.35cm \chi = az \sac \phi=\phi(u) \,,
\label{sol2}
\eea
where $\cf, \cb$ and $\ch$ are functions of the holographic radial coordinate $u$. Reparametrization invariance has been used to fix $g_{xx}$ and $g_{yy}$ in \eqn{sol1}; once this is done, $\cb$ cannot be set to unity in general. The solution is clearly isotropic in the $xy$-directions, but not in the $z$-direction, unless $\ch=1$. The axion, which is dual to the gauge theory $\theta$-term, is responsible for inducing the anisotropy. $\cf$ is a `blackening factor' that will vanish at the position of the horizon, $u=\uh$. The boundary of the spacetime is at $u=0$. 

The functions $\cf, \cb$ and $\ch$ only depend on the radial coordinate $u$ and are given in terms of the dilaton as 
\bea
\ch&=& e^{-\phi} \,, \label{eq_H} \\
\cf&=& \frac{e^{-\frac{1}{2}\phi}}{4(\phi'+u\phi'')}\left[ a^2 e^{\frac{7}{2}\phi}(4u+u^2\phi')+16\phi'\right]\,,
\label{eq_F} \\
\frac{\cb'}{\cb}&=& \frac{1}{24+10 u\phi'}\left(24\phi'-9u\phi'^2+20u\phi''\right)\,,
\label{eq_B}
\eea
where the primes denote differentiation with respect to $u$. Eqn.~\eqn{eq_B} determines $\cb$ up to an arbitrary multiplicative constant.  Eqns.~\eqn{sol1}-\eqn{eq_B} are a solution of the field equations provided the dilaton obeys the third-order equation
\bea
0&=&\frac{256 \phi ' \phi ''-16 \phi '^3
   \left(7 u \phi '+32\right)}{u \, a ^2 e^{\frac{7 \phi }{2}}
   \left(u \phi '+4\right)+16 \phi '} +\frac{\phi ' }{u \left(5 u \phi '+12\right) \left(u \phi
   ''+\phi '\right)} \times \nonumber \\
   && \times \Big[13 u^3 \phi '^4+8 u \left(11
   u^2 \phi ''^2-60\phi''-12 u \phi ''' \right)  
   +u^2 \phi '^3
   \left(13 u^2 \phi ''+96\right) \nonumber \\
   && +2 u \phi '^2 \left(-5 u^3 \phi
   '''+53 u^2 \phi ''+36\right)+\phi ' \left(30 u^4 \phi
   ''^2-64 u^3 \phi '''-288+32 u^2 \phi
   ''\right) \Big] \,.
   \label{eq_dil}
\eea

Without loss of generality (see Section \ref{number}), we will seek solutions such that the dilaton vanishes at the boundary, i.e.~such that $\phib\equiv \phi(0)=0$.
Eqs.~\eqn{eq_H} and \eqn{eq_F} expanded near the boundary then imply $\Fb=\Hb=1$. 
We will also choose the arbitrary constant in $\cb$ such that $\Bb=1$. 
With these normalizations the temperature and the entropy density are readily computed. We first note that the Euclidean continuation of the metric \eqn{sol1} in the $(t_\mt{E},u)$-plane near $\uh$ reads
\bea
ds^2_\mt{E} \approx \frac{e^{-\frac{1}{2}\phih}}{\uh^2} \, 
\Big[ \cf_1 \cb_\mt{H} (\uh-u)\, dt_\mt{E}^2 + \frac{du^2}{\cf_1 (\uh-u)} \Big] \,,
\label{euclideanized}
\eea
where $\cf_1=-\cf'(\uh)$. The standard requirement that the metric be regular at $u=\uh$ then determines the period of the Euclidean time, $\delta t_\mt{E}$, which we identify with the inverse temperature:
\bea
T = \frac{1}{\delta t_\mt{E}}  =  \frac{\cf_1 \sqrt{\cb_\mt{H}}}{4\pi} \,.
\label{temperature}
\eea
The entropy density is simply obtained from the area of the horizon. The area element on a $t=\mbox{const.}, u=\uh$ hypersurface is 
\bea
dA_\mt{H} = \frac{e^{-\frac{5}{4}\phi_\mt{H}}}{\uh^{3}} \, dx\, dy\, dz  \,,
\nonumber
\eea
so the entropy density per unit volume in the $xyz$-directions is
\be
s = \frac{A_\mt{H}}{4 G V_3} = 
\frac{\pi^2}{2}\nc^2 \times 
\frac{e^{-\frac{5}{4} \phih}}{\pi^3 \uh^3} \,.
\label{entropy}
\ee
The isotropic black D3-brane solution is a solution of the equations above with $a=0$ and  
\be 
\phi = 0 \sac \cb=\ch=1 \sac \cf=1-\frac{u^4}{\uh^4} \sac \uh=\frac{1}{\pi T} \,.
\label{isotropic}
\ee
Eqn.~\eqn{entropy} then yields the familiar expression for the entropy density of 
${\cal N}=4$ SYM \cite{correct}:
\be
s^0 (T)=\frac{\pi^2}{2} \nc^2 \, T^3 \,.
\label{entropyN=4}
\ee

\begin{figure}
\begin{center}
\includegraphics[scale=0.9]{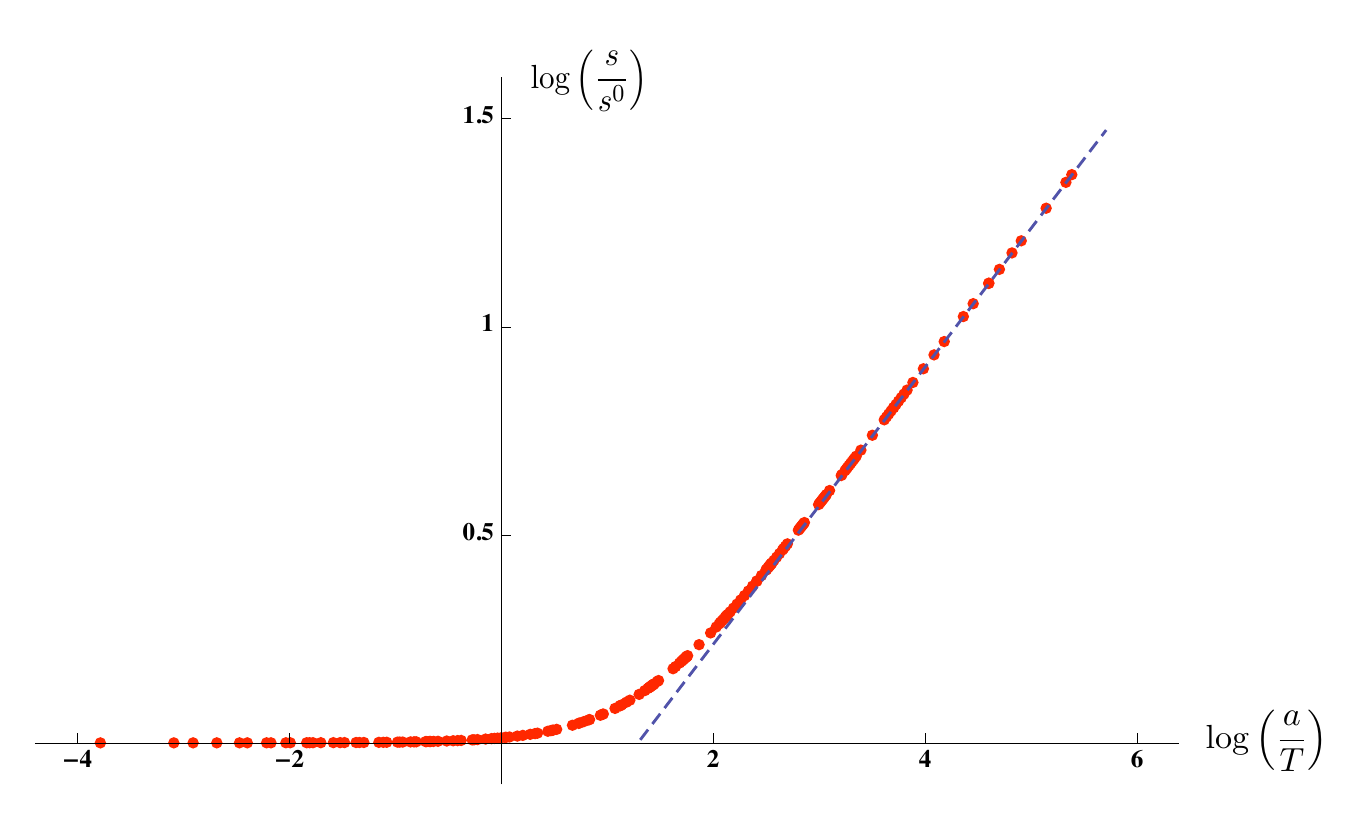}
\caption{\small Log-log plot of the entropy density as a function of $a/T$, with $s^0$ defined as in eqn.~\eqn{entropyN=4}. The dashed blue line is a straight line with slope $1/3$.
\label{scalings}}
\end{center}
\end{figure}

At this point we can perform an interesting check on our solution. As mentioned above, the zero-temperature solution is a domain-wall-like solution in the radial direction, interpolating between an AdS geometry in the UV and a Lifshitz-like geometry in the IR \cite{ALT}. The radial position at which the transition takes place is set by the anisotropic scale, $a$. Thus we expect that in the limit $T\gg a$ the entropy density should scale as in \eqn{entropyN=4}, since in this limit the horizon should lie in the asymptotic region where the geometry is approximately AdS. In the opposite limit we expect the entropy density to scale as 
\be
s = c_\mt{ent} \nc^2 a^{1/3} T^{8/3} 
\label{slif}
\ee
with $c_\mt{ent}$ a numerical coefficient, since this is the appropriate scaling \cite{ALT} in the IR Lifshitz-like region in which the horizon lies when $T\ll a$ (see Appendix~\ref{comparison} for details). These scalings are precisely reproduced by the entropy density computed with our numerical solution (see Appendix \ref{derivation} for details), as shown in Fig.~\ref{scalings}. 
This plot was produced by evaluating the entropy density for many different values of $a$ and $T$. We see that for $a\ll T$ the points are aligned along the horizontal axis, thus reproducing (\ref{entropyN=4}). In the opposite regime $a \gg T$, the points are aligned instead along a straight line with slope $1/3$, which means that the entropy density scales in this case as in \eqn{slif}. In between, the entropy density smoothly interpolates between the two limiting behaviours. 


\section{Stress tensor}
\label{stress-section}

The energy density and the pressures of the deformed $\cn=4$ theory can be obtained by calculating the holographic stress tensor, defined as the variation of the supergravity action with respect to the boundary metric. The action suffers from large-volume divergences, which can be regularized and subtracted by a procedure called holographic renormalization (see e.g.~\cite{Skenderis:2002wp} and references therein). The total on-shell action takes the form 
\be
S_\mt{on-shell}=S_\mt{bulk}-S_\mt{ct} \,,
\label{on}
\ee
where  the counterterm action can be obtained by particularizing the results of Refs.~\cite{odin,Yiannis} to the case of interest here. Since we will be interested in the thermodynamics of the system it is convenient to work here in Euclidean signature. In this case $S_\mt{bulk}$ is given by the Euclidean continuation of \eqn{action} and, omitting terms that vanish identically for the case at hand, the counterterm action takes the form 
\be
S_\mt{ct}=\frac{1}{\kappa^2} \int d^4x \sqrt{\g}
\left( 3- \frac{1}{8} e^{2\phi}\pa_i\chi\pa^i\chi \right) 
- \log v  \int d^4x \sqrt{\g} {\cal A} + \frac{1}{4}(\cs-1) \int d^4x \sqrt{\g} {\cal A} \,,
\label{counterterms}
\ee
where ${\cal A}(\g_{ij},\phi,\chi)$ is the conformal anomaly in the axion-dilaton-gravity system \cite{odin,Yiannis}.
The coordinate $v$ is the standard Fefferman-Graham (FG) coordinate, in terms of which the metric near the boundary takes the form
\be
ds^2=\frac{dv^2}{v^2} + \gamma_{ij}(x,v)\,  dx^i dx^j \,.
\label{FG}
\ee 
Note that if we were to restore the AdS radius $L$ then the coefficient of the second term in eqn.~\eqn{counterterms} would become $\log(v/L)$.

The counterterm action \eqn{counterterms} is understood to be evaluated on a constant-$v$ slice, the bulk action is understood to be evaluated by integrating from the horizon down to the lower cut-off $v$, and the limit $v\ra 0$ is understood to be taken at the end of the calculation. The logarithmic term in \eqn{counterterms} is necessary to cancel certain divergences from the bulk action, and it breaks diffeomorphism invariance in the bulk. This is the origin of the conformal anomaly in the boundary. The third term in \eqn{counterterms} is finite and can therefore be added with an arbitrary coefficient, which we have written as $(\cs-1)/4$ to simplify subsequent equations. The freedom to add this term is part of the general freedom in the choice of renormalization scheme.

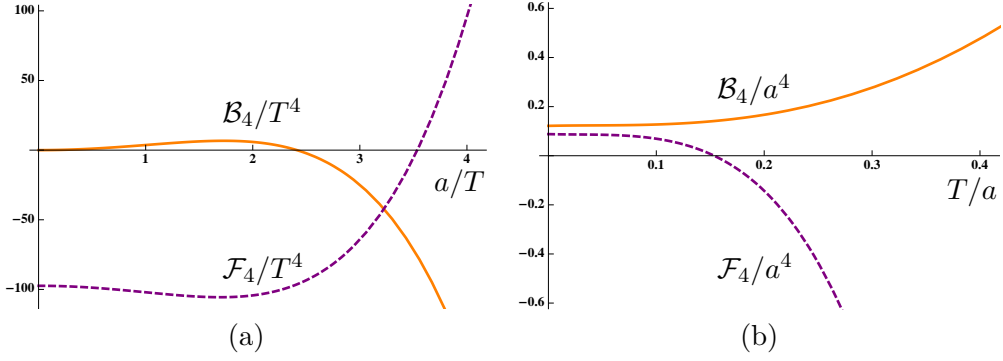
\begin{figure}[ht]
\begin{center}
\begin{tabular}{cc}
\includegraphics[scale=0.7]{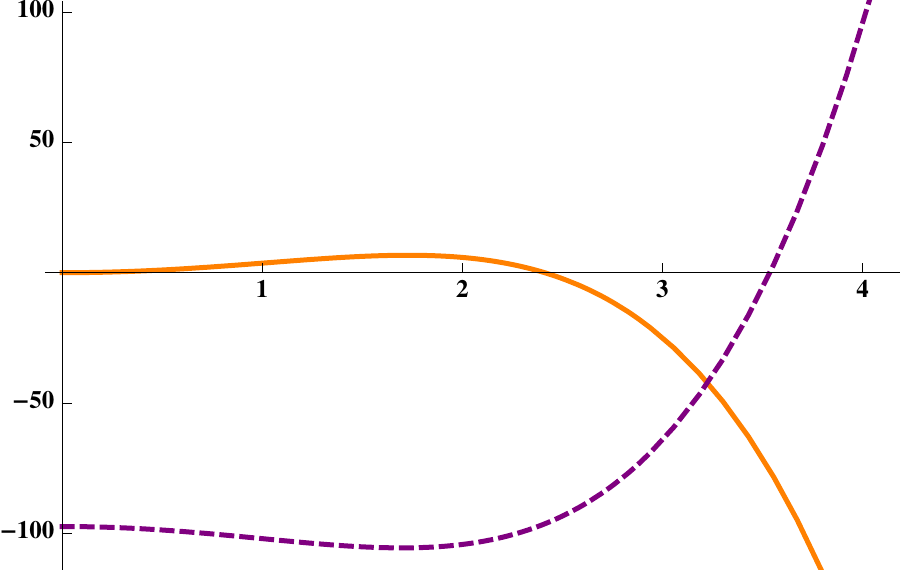}
\put(-20,45){$a/T$}
\put(-100,71){$\cb_4/T^4$}
\put(-100,12){$\cf_4/T^4$}
&
\includegraphics[scale=0.7]{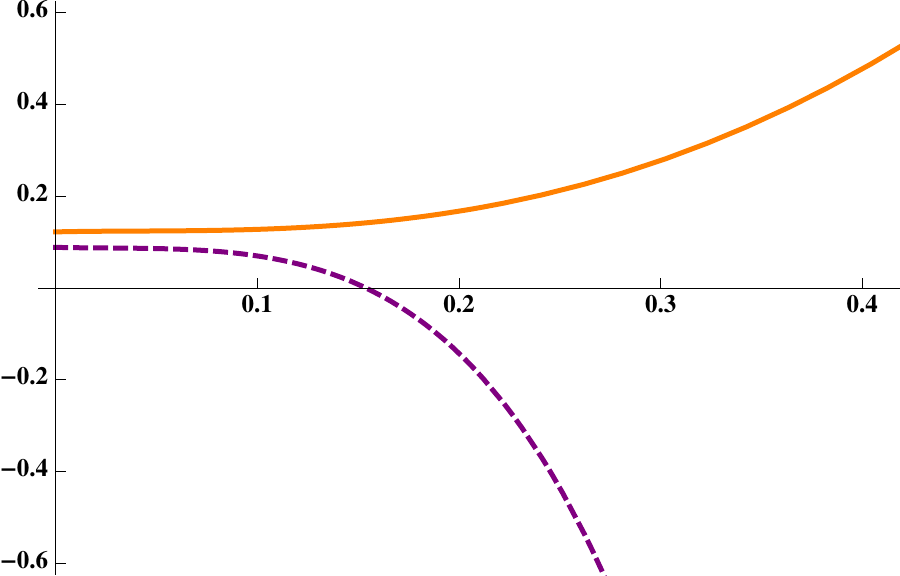}
\put(-20,42){$T/a$}
\put(-107,80){$\cb_4/a^4$}
\put(-107,12){$\cf_4/a^4$}
\\
(a) & (b)
\end{tabular}
\caption{\small Expansion coefficients ${\cal F}_4$  and 
${\cal B}_4$. In (a) we fix $T= 1$ and plot $\cf_4/T^4$ and $\cb_4/T^4$ as functions of $a/T$. In (b) we fix $a= 1$ and plot $\cf_4/a^4$ and $\cb_4/a^4$
as functions of $T/a$. From the plots in (a) we can see that, as $a\to 0$, we recover the isotropic solution (\ref{isotropic}), since in this limit ${\cal B}_4\to 0$ and ${\cal F}_4\to -97.4 \simeq -\pi^4$.
\label{f4b4}
}
\end{center}
\end{figure}

The fields have the following general expansions near the boundary: 
\bea
\gamma_{ij} &=& \frac{1}{v^2} \left[ g^{(0)}_{ij} + v^2 g^{(2)}_{ij} 
+ v^4 \left( g^{(4)}_{ij} + 2 \log v \, \tilde{g}^{(4)}_{ij} \right) + {O}(v^6) \right] 
\label{gammaExpansion}\,, \\
\phi &=& \phi^{(0)} + v^2 \phi^{(2)}
+ v^4 \left( \phi^{(4)} + 2 \log v \, \tilde{\phi}^{(4)} \right) +  {O}(v^6) \,, \\
\chi &=& \chi^{(0)} + v^2 \chi^{(2)}
+ v^4 \left( \chi^{(4)} + 2 \log v \, \tilde{\chi}^{(4)} \right) +  {O}(v^6) \,.
\eea
In our case, $g^{(0)}_{ij}=\eta_{ij}$ is the flat gauge theory metric. The axion is simply 
\be
\chi=  \chi^{(0)}=az\,, 
\ee
and the dilaton takes the form
\be
\phi = -\frac{a^2}{4}v^2+\left(\frac{2\, {\cal B}_4}{7}-\frac{47\, a^4}{4032}\right)v^4-\frac{a^4}{6}v^4 \log v +  {O}(v^6) \,. 
\ee
The expansion of the metric can be obtained from the expansion of the functions $\cf, \cb$ and $\ch$,
\bea
&& \cf = 1+\frac{11\, a^2}{24}v^2+\left({\cal F}_4+\frac{11\, a^4}{144}\right) v^4+\frac{7\, a^4}{12}v^4 \log v +  {O}(v^6) \,, \nonumber \\
&& \cb = 1-\frac{11\, a^2}{24}v^2+\left({\cal B}_4-\frac{11\, a^4}{144}\right)v^4-\frac{7\, a^4}{12}v^4 \log v +  {O}(v^6) \,, \nonumber \\
\cr && \ch =  1+\frac{a^2}{4}v^2-\left(\frac{2\, {\cal B}_4}{7}-\frac{173\, a^4}{4032}\right)v^4+\frac{a^4}{6}v^4 \log v +  {O}(v^6) \,,
\label{expansionFunctions}
\eea
and it reads 
\bea
g_{tt} &=& -1 + \frac{a^2}{24} v^2 + \left(
- \frac{23 {\cal B}_4}{28}- \frac{3 {\cal F}_4}{4} + \frac{2749 a^4}{16128} 
+ \frac{a^4}{48} \, 2 \log v \right) v^4+  {O}(v^6)  \,,
 \nonumber \\ 
g_{xx} = g_{yy} &=& 1 - \frac{a^2}{24} v^2 + \left(
- \frac{5 {\cal B}_4}{28}- \frac{ {\cal F}_4}{4} + \frac{71 a^4}{1792} 
- \frac{a^4}{48} \, 2 \log v \right) v^4+  {O}(v^6)  \,,
\nonumber \\ 
g_{zz} &=& 1 + \frac{5 a^2}{24} v^2 + \left(
- \frac{13 {\cal B}_4}{28}- \frac{ {\cal F}_4}{4} + \frac{1163 a^4}{16128} 
+ \frac{a^4}{16} \, 2 \log v \right) v^4+  {O}(v^6)  \,.
\label{metricExpansion}
\eea
${\cal F}_4 (a,T)$ and ${\cal B}_4(a,T)$ are two coefficients that are not determined by the asymptotic equations of motion, as expected, but that can be read off once a particular solution is known. Below we will obtain their expressions analytically for large and small $T/a$. They can also be computed for arbitrary values of $T/a$ by fitting the asymptotic fall-off of the numerical solution, and the result is shown in Fig. \ref{f4b4}. 
These coefficients enter the relation between the FG coordinate $v$ and the $u$ coordinate of \eqn{sol1}-\eqn{sol2}, which takes the form 
\bea
v &=& u -  \frac{a^2}{12} u^3 + \left( - \frac{{\cal B}_4 + 7 {\cal F}_4}{56} 
+ \frac{1091 a^4}{32256} 
- \frac{a^4}{16} \log u \right) u^5 +  {O}(u^7) 
\,, \label{vOFu} \\
u &=& v + \frac{a^2}{12} v^3 + \left( \frac{{\cal B}_4 + 7 {\cal F}_4}{56} 
- \frac{419 a^4}{32256} 
+ \frac{a^4}{16} \log v \right) v^5 +  {O}(v^7) 
 \,.
\label{uOFv}
\eea
In fact, $\cf_4$ and $\cb_4$ are the coefficients of the $ {O}(u^4)$ terms in the near-boundary expansion of the functions $\cf$ and $\cb$ in terms of the $u$-coordinate. The change to the FG coordinate $v$ introduces the additional $a^4$ terms in the $ {O}(v^4)$ coefficients in eqn.~\eqn{expansionFunctions}.

Note that the logarithmic terms in \eqn{metricExpansion} are independent of $\cf_4$ and $\cb_4$ and are given by 
\be
\tilde{g}^{(4)}_{ij} = \frac{a^4}{48} \, h_{ij} \sac
h_{ij} = \mbox{diag} \left(1,-1,-1,3 \right) \,.
\label{hij}
\ee
Observe too that  $\tilde{g}^{(4)}_{ij}$ is traceless, 
i.e.~$\mbox{Tr} \, \tilde{g}^{(4)} = {g}_{(0)}^{ij} \tilde{g}^{(4)}_{ij}=0$. 

Again using the results from \cite{Yiannis}, the expectation value of the stress tensor is found to be 
\be
\vev{T_{ij}} = \frac{2}{\kappa^2} \left[
g^{(4)}_{ij} - \left( \mbox{Tr} \, g^{(4)} \right) g^{(0)}_{ij} 
+ \frac{\cs}{2} \, \tilde{g}^{(4)}_{ij} + \cdots \right] \,,   
\label{tensor}
\ee
where the dots stand for terms that do not depend on $g^{(4)}_{ij}, \phi^{(4)}$ or 
$\chi^{(4)}$. When evaluated on the solution \eqn{metricExpansion} this becomes
\be
\vev{T_{ij}} = \mbox{diag}(E, P_{xy}, P_{xy}, P_z)\,,
\ee
with
\bea
E&=& \frac{\nc^2 }{2\pi^2}\left(-\frac{3}{4}{\cal F}_4 -
\frac{23}{28}{\cal B}_4
+\frac{2777}{16128}a^4+\frac{c_\mt{sch}}{96}a^4\right) ,\nonumber \\
P_{xy}&=& 
\frac{ \nc^2 }{2\pi^2}\left(-\frac{1}{4}{\cal F}_4 -\frac{5}{28}{\cal B}_4+ \frac{611}{16128}a^4-\frac{c_\mt{sch}}{96}a^4\right) ,\nonumber \\
P_z&=&
\frac{ \nc^2 }{2\pi^2}\left(-\frac{1}{4}{\cal F}_4 
-\frac{13}{28}{\cal B}_4
+\frac{2227}{16128}a^4+\frac{c_\mt{sch}}{32}a^4\right) .
\label{stress_tensor}
\eea
We use the notation $P_{xy}\equiv P_x=P_y$ as a reminder that the pressures in the $x$- and $y$-directions are equal. Similar expressions can be obtained \cite{Yiannis} for the expectation values of the operators dual to $\phi$,
${\cal O}_\phi \sim \mbox{Tr} \, F^2$,  and to $\chi$, 
${\cal O}_\chi \sim \mbox{Tr} \, F\tilde F$, from which one finds that $\langle {\cal O}_\phi \rangle \neq 0$ and $\langle {\cal O}_\chi \rangle =0$.  The near-boundary expansion determines the divergence and the trace of the stress tensor, which are independent of the ${\cal F}_4$ and ${\cal B}_4$ coefficients. Specifically, the divergence obeys
\be
\partial^i \vev{T_{ij}} 
+ \langle {\cal O}_\phi \rangle \partial_j  \phi^{(0)}
+ \langle {\cal O}_\chi \rangle \partial_j  \chi^{(0)} = 0 \,.
\ee
In our case this immediately implies the conservation of the stress tensor, 
$\partial^i \vev{T_{ij}} =0$, by virtue of the facts that $\partial_j  \phi^{(0)}=0$ and $\langle {\cal O}_\chi \rangle =0$. 
The former is simply due to the fact that there is no source for 
${\cal O}_\phi$ in our solution, namely the fact that the non-normalizable mode of the dilaton vanishes, $\phi^{(0)}=0$. The latter is due to the fact that, although there is a source for the instanton density ${\cal O}_\chi$ because the non-normalizable mode of the axion is non-zero, $ \chi^{(0)} \neq 0$, this does not induce an expectation value for  ${\cal O}_\chi$.
We thus confirm  that translation invariance is preserved, as anticipated above. Similarly, the trace obeys 
\be
\vev{T_i^i} = {\cal A}(g^{(0)}_{ij}, \phi^{(0)}, \chi^{(0)})  \,,
\label{trace}
\ee
where 
\be  
{\cal A}(g^{(0)}_{ij}, \phi^{(0)}, \chi^{(0)})= 
\frac{1}{12 \kappa^2} \left(g_{(0)}^{ij} \pa_i \chi^{(0)} \pa_j \chi^{(0))} \right)^2
= \lim_{v\ra 0} \sqrt{\g} \, {\cal A}(\gamma_{ij},\phi,\chi) =
\frac{\nc^2 a^4}{48 \pi^2}
\label{anomaly}
\ee
is the conformal anomaly evaluated on our specific solution. 
This type of `matter conformal anomaly' is related to short-distance singularities in the two-point function of ${\cal O}_\chi$ and was first studied in \cite{kostasNOrenorm}, where it was shown that it does not renormalize in the sense that its coefficient is fixed relative to the normalization of the two-point function. On the gravity side the anomaly was first computed in \cite{kostasgravity} and is directly related to the fact that some of the coefficients of the logarithmic terms in the expansion above, concretely $\tilde{g}^{(4)}_{ij}$ and $ \tilde{\phi}^{(4)}$, do not vanish. We will see that the anomaly has important consequences for the thermodynamics of our system. 

For our purposes it is crucial to understand how the terms in the near-boundary expansion transform under a rescaling of the coordinates of the form 
\be
x_i = k \,  x_i' \sac v =  k v' \,,
\label{resc}
\ee
where $k$ is a positive real number.
In the gauge theory this is equivalent to the rescaling of $a$ and $T$ 
\be
a' = k a \sac T' = k T \,,
\label{aT}
\ee
as is easily verified from their definitions in terms of parameters on the gravity side:
\be
\chi = a z = (k a) z' \sac 
(k T) = \frac{k}{\delta t_\mt{E}} = \frac{1}{\delta t_\mt{E}'} \,.
\ee
Following \cite{kostas}, we note that the rescaling above leaves the FG form of the metric \eqn{FG} invariant and transforms all the expansion coefficients homogeneously, 
\be
g_{(0)  ij}' =  g_{(0)  ij} \sac 
g_{(2)  ij}' = k^2 g_{(2)  ij} \sac
\tilde g_{(4)  ij}' = k^4 \tilde g_{(4)  ij}  \,,
\ee
except for $g_{(4)  ij}$, which acquires an inhomogeneous piece due to the logarithmic term in \eqn{gammaExpansion}:
\be
 g_{(4)  ij}' = k^4 g_{(4)  ij} + 
 2\, k^4 \log k \, \tilde g_{(4)  ij} \,.
\ee
An analogous statement holds for the transformation properties of the expansion coefficients of $\phi$ and $\chi$. It follows that the stress tensor \eqn{tensor} transforms as 
\be
\vev{T_{ij}'} = k^4 \, \vev{T_{ij}} +  
\frac{4}{\kappa^2}  k^4 \log k \, \tilde g_{(4)  ij} \,,
\ee
namely
\be
\vev{T_{ij} (k a, k T)} = k^4 \, \vev{T_{ij} (a,T)} +  
k^4 \log k \, \frac{\nc^2 a^4}{48\pi^2} \, h_{ij} \,,
\label{inhom}
\ee
where we have made use of \eqn{hij}. This immediately implies that the stress tensor must take the form
\be
\vev{ T_{ij} (a,T)} = a^4  \, t_{ij} \left( \frac{a}{T} \right) 
 + a^4 \log \left(\frac{a}{\mu} \right) \, \frac{\nc^2}{48\pi^2} \, h_{ij} \,,
 \label{later}
\ee
where $\mu$ is some arbitrary reference scale, a remnant of the renormalization process much like the subtraction point in QCD. The first and second terms transform homogeneously and inhomogeneously under the rescaling \eqn{aT}, respectively. Needless to say, one could rewrite the first term in a variety of forms, e.g.~$T^4 t_{ij}(a/T)$, etc. Also, one could replace $\log (a/\mu)$ by $\log (a/T) + \log (T/\mu)$, thus redefining 
\be
t_{ij} \ra t_{ij} + \frac{\nc^2}{48 \pi^2} \log \left( \frac{a}{T} \right) h_{ij} \,.
\ee
In any case, the key conclusion is that the physics does not only depend on the ratio $a/T$, but on the \emph{two} independent dimensionless ratios that can be built from $a, T$ and $\mu$.

We now realize that the freedom in the choice of scheme associated to shifts of $\cs$ in the counterterm action \eqn{counterterms} is precisely equivalent to a rescaling of the reference scale $\mu$. This follows from the fact that, as we see from eqs.~\eqn{tensor}, \eqn{hij} and \eqn{later}, both transformations shift the stress tensor by an amount proportional to ${\cal A} \times h_{ij}$. The fact that one must introduce a reference scale $\mu$ in order to define the theory with $a \neq 0$ is a direct consequence of the conformal anomaly. There is a (limited) analogy with the situation in QCD with one quark flavour with $\mq\neq 0$. The classical QCD theory has only one scale, $\mq$. However, upon renormalization one realizes that a scale $\mu$ must be introduced  in order to define the quantum theory. This scale is arbitrary, and one can trade it, for example, for the scale $\Lambda_\mt{QCD}$ at which the coupling constant becomes strong. Similarly, in the present case the classical theory (at zero temperature) has only one scale, $a$, but the quantum theory depends on $a$ and $\mu$. 
The reason this analogy is not exact is that the scale $\Lambda_\mt{QCD}$ in QCD survives even in the limit $\mq=0$, whereas in our case the theory with $a=0$ is truly conformal: in this limit there is no anomaly and the scale $\mu$ disappears. Note also that, if we were interested in working with the theory at a fixed value of $a$, we could choose $\mu=a$ thereby seemingly  eliminating  the $\mu$-dependence all together. However, we will be interested in examining how the physics depends on $a$, in which case the choice above is not appropriate. For example, in taking derivatives of the free energy with respect to $a$ to obtain the chemical potential, as in \eqn{sphi2}, $a$ and $\mu$ must be treated as independent quantities. 

Below we will obtain analytical expressions for the energy and the pressures \eqn{stress_tensor} in the limits of large and small $T/a$. For arbitrary $T/a$ their values can be obtained numerically. The results for some specific cases are shown in Figs.~\ref{enpress_aoverT} and \ref{enpress_Tovera}. In Fig.~\ref{enpress_aoverT}  we fix $T\simeq 0.33$ and $T\simeq 1.1$ and normalize the results by the isotropic $\cn=4$ SYM values 
\be
E^0 (T) = \frac{3\pi^2 \nc^2 T^4}{8} \sac
P^0 (T) = \frac{\pi^2 \nc^2 T^4}{8} \,.
\label{EP0}
\ee
We see from Fig.~\ref{enpress_aoverT} that in the limit $a\ra 0$ the energy and the pressures approach their isotropic values, as expected.
In Fig.~\ref{enpress_Tovera}   we fix $a\simeq 0.34$ and $a\simeq 2.86$ and normalize the energy and the pressures by $\nc^2 a^4$. Note that the energy and/or the pressures can become negative at low temperatures depending on the value of $a/\mu$, as is in fact apparent from \eqn{later}. We will come back to the interpretation of these results below. 
\begin{figure}[tb]
\begin{center}
\begin{tabular}{cc}
\includegraphics[scale=0.7]{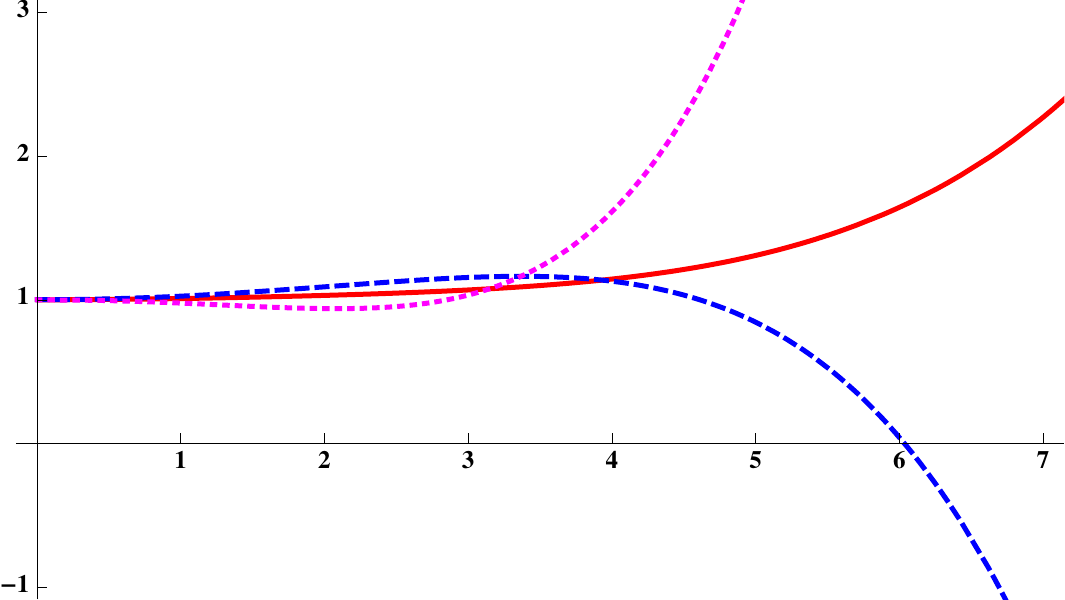}
\put(-20,37){$a/T$}
\put(-45,90){$E/E^0$}
\put(-103,110){$P_{z}/P^0$}
\put(-59,3){$P_{xy}/P^0$}
&
\includegraphics[scale=0.7]{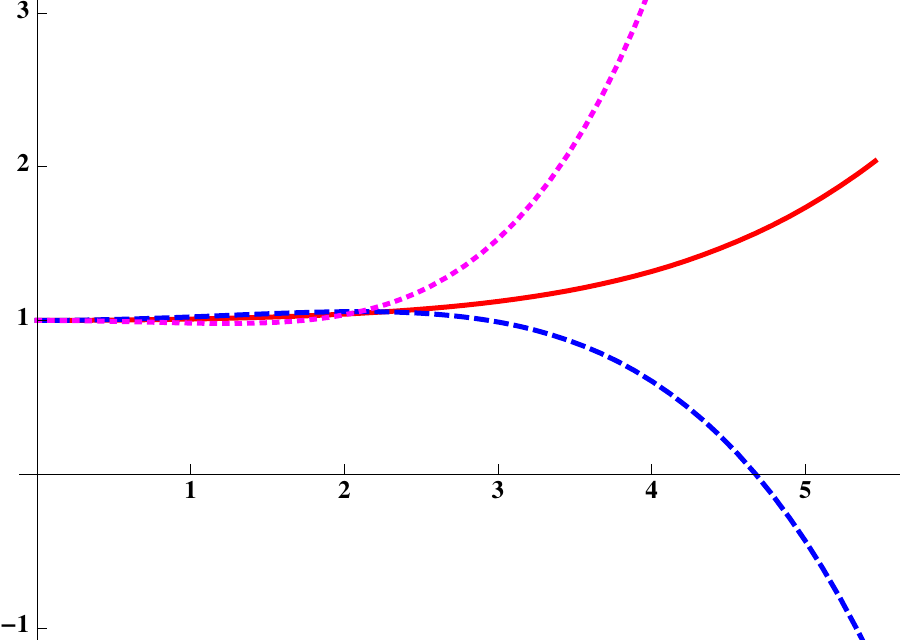}
\put(-20,40){$a/T$}
\put(-45,90){$E/E^0$}
\put(-93,110){$P_{z}/P^0$}
\put(-55,3){$P_{xy}/P^0$}
\\
(a) & (b)
\end{tabular}
\caption{\small The energy and pressures normalized by their isotropic values \eqn{EP0} as functions of $a/T$, with $T\simeq 0.33$ in (a) and $T\simeq 1.1$ in (b). We have chosen $\cs=-1$ and $\mu=1$. 
}
\label{enpress_aoverT}
\end{center}
\end{figure}
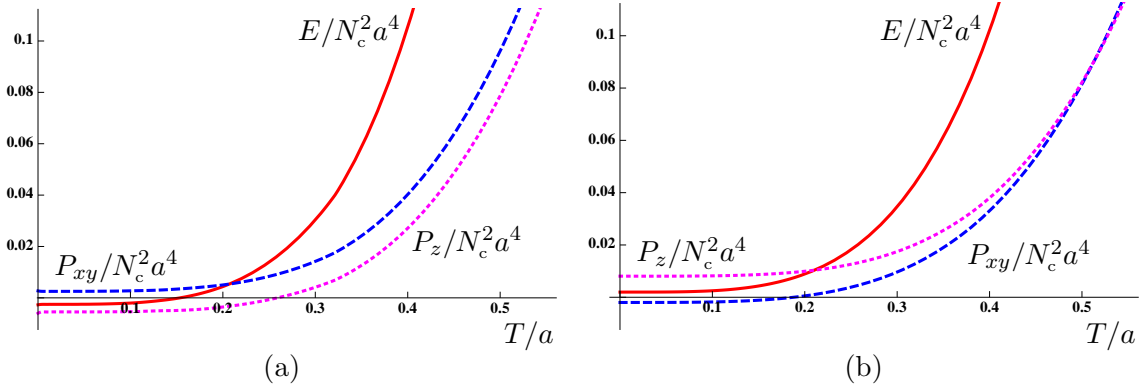
\begin{figure}[tb]
\begin{center}
\begin{tabular}{cc}
\includegraphics[scale=0.8]{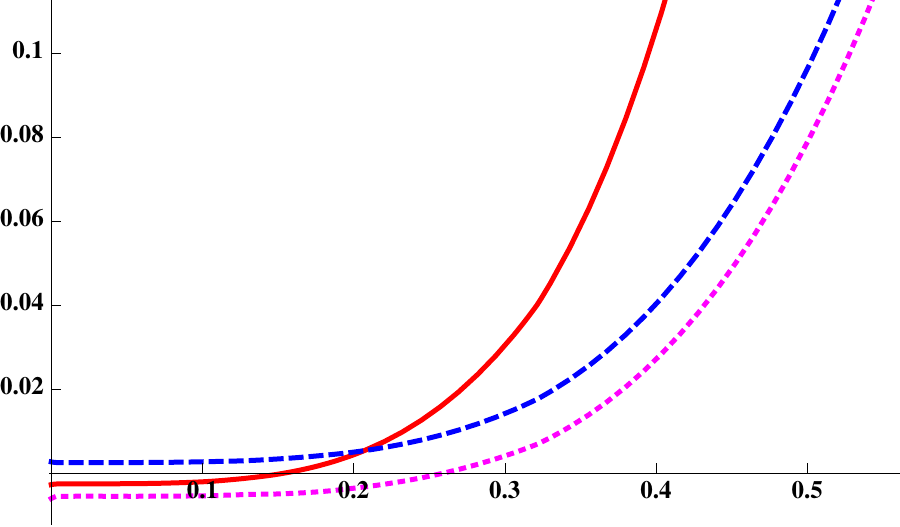}
\put(-20,-5){$T/a$}
\put(-190,21){$P_{xy}/\nc^2 a^4$}
\put(-98,110){$E/\nc^2 a^4$}
\put(-55,30){$P_z/\nc^2 a^4$}
&
\includegraphics[scale=0.8]{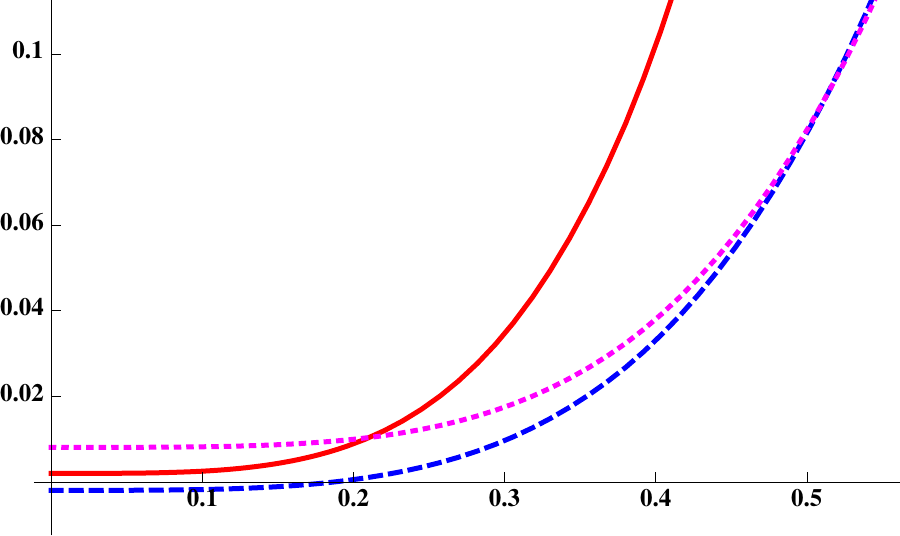}
\put(-20,-5){$T/a$}
\put(-190,26){$P_{z}/\nc^2 a^4$}
\put(-98,110){$E/\nc^2 a^4$}
\put(-65,25){$P_{xy}/\nc^2 a^4$}
\\
(a) & (b)
\end{tabular}
\caption{\small The energy and pressures divided by $\nc^2 a^4$ as functions of $T/a$, with $a \simeq 0.34$ in (a) and $a\simeq 2.86$ in (b). We have chosen $\cs=-1$ and $\mu=1$. 
}
\label{enpress_Tovera}
\end{center}
\end{figure}


\section{Thermodynamics}
\label{sec-thermodynamics}

\subsection{Number of D7-branes}
\label{number}
\begin{figure}
\begin{center}
\includegraphics[scale=0.8]{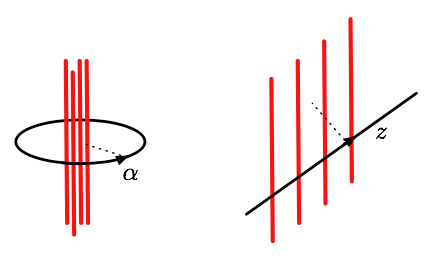}
\caption{\small Distributions of D7-branes.
\label{circulation}}
\end{center}
\end{figure}

As anticipated in the Introduction, the solution at hand can be thought of as describing a uniform D7-brane charge density per unit length in the $z$-direction \cite{ALT}. To obtain the correct normalization, a reminder about conventions is necessary. The ten-dimensional supergravity action in the string frame takes the form 
\be
S_\mt{sugra} = \frac{1}{2\tilde \kappa_\mt{10}^2} \int \sqrt{-g} \left( e^{-2\varphi} R - \sum_n  \frac{1}{2 \cdot n!} \,
\tilde F_n^2 + \cdots \right) \,,
\ee
where $2\tilde \kappa_\mt{10}^2 = (2\pi)^7 \ell_s^8$ and $\tilde F_n$ denote the Ramond-Ramond (RR) field strengths of different degrees. In particular, $\tilde F_1 = d\tilde \chi$. (The reason for the tildes will be clear shortly.) The dilaton $\varphi$ appearing in this equation is \emph{not} normalized to zero at the boundary, but in fact determines the string and the SYM coupling constants as \mbox{$\gym^2/4\pi=g_s=e^{\varphi_\mt{bdry}}$}. Similarly, the theta-angle is related to the boundary value of $\tilde \chi$ as $\theta = 2\pi \tilde \chi_\mt{bdry}$. 
Given a collection of overlapping D7-branes (see Fig.~\ref{circulation}(left)), their total number is measured by the circulation of the axion around the branes, 
\be
N_\mt{D7} = \frac{1}{2\tilde \kappa^2 \tilde T_\mt{D7}} \int  d \tilde \chi \,,
\ee
as appropriate since D7-branes are magnetically charged with respect to the axion. In this equation, $\tilde T_\mt{D7}^{-1} = (2\pi)^7 \ell_s^8$ is the inverse D7-brane tension, and we note that $2\tilde \kappa^2 \tilde T_\mt{D7}=1$. By symmetry, in this case we have  
$\tilde \chi =\alpha N_\mt{D7}/2\pi$, with $\alpha$ the polar angle in the plane transverse to the branes. Suppose now that we distribute the D7-branes homogeneously along the $z$-direction, as in Fig.~\ref{circulation}(right), then a similar symmetry argument shows that in this case $\tilde \chi = n_\mt{D7} z$, where 
$n_\mt{D7} = N_\mt{D7} /L_z$ is the density of D7-branes along the $z$-direction.\footnote{Unless otherwise stated, in this paper we take $N_\mt{D7},\, L_z\to \infty$ while maintaining $n_\mt{D7}$ finite, i.e. the $z$-direction is non-compact. One could consider instead the $z$-coordinate to be periodically identified with period $L_z< \infty$. This was the choice adopted in \cite{QHE}. The fact that the action (\ref{def}) (with anti-periodic boundary conditions in the $z$-direction for the fermions) flows in the IR to a 3-dimensional pure Chern-Simons theory was used in that reference to holographically model the fractional quantum Hall effect. }
It follows that the $\theta$-angle is given by 
\be
\theta=2\pi n_\mt{D7} z \,.
\label{theta}
\ee

In this paper we found it convenient to work with a dilaton $\phi$ normalized in such a way that $\phi_\mt{bdry}=0$. This can be achieved by defining
\be
e^\phi = g_s e^\varphi  \sac F_n = g_s \tilde F_n \,.
\label{res}
\ee
Under this rescaling the supergravity action rescales homogeneously with a factor of $g_s^{-2}$ and takes the form 
\be
S_\mt{sugra} = \frac{1}{2 \kappa_\mt{10}^2} \int \sqrt{-g} \left( e^{-2\phi} R 
- \sum_n  \frac{1}{2 \cdot n!} \, F_n^2 + \cdots \right) \,,
\ee
where now $2\kappa_\mt{10}^2 = (2\pi)^7 \ell_s^8 g_s^2$. The fields appearing in this action are the ones we have used throughout this paper.
Similarly, the D7-brane action transforms as 
\bea
S_\mt{D7} &=& - \tilde T_\mt{D7} \int  e^{-\varphi} \sqrt{-g} + \cdots \nonumber \\
&=&  - T_\mt{D7} \int  e^{-\phi} \sqrt{-g} + \cdots \,,
\eea
and therefore $T_\mt{D7}^{-1} = (2\pi)^7 \ell_s^8 g_s$. We see that the coefficient $a$ in the rescaled axion $\chi = az$ is given by
\be
a=g_s n_\mt{D7} = \frac{\lambda}{4\pi} \frac{n_\mt{D7}}{\nc}  \,,
\label{a}
\ee
where $\lambda=\gym^2 \nc=4\pi g_s \nc$ is the gauge theory `t Hooft coupling. Interestingly, the factor on the right hand side is analogous to the one found \cite{back,back2,back3}\footnote{Note that the convention used in \cite{back,back2} was $\gym^2=2\pi g_s$.} to control the backreaction of flavour D7-branes on the geometry dual to ${\cal N}=4$ SYM plasma with quarks \cite{flavour}. Note that in this paper we keep $a$ fixed as $\nc \ra \infty$ in order to obtain a finite limit, which implies that the $\theta$-angle and the D7-brane density scale as 
\be
\theta \sim \n \sim \frac{\nc}{\lambda} \,.
\label{thetan}
\ee
The power of $\nc$ is expected from the fact that $\theta/\nc$ should be kept fixed in order to achieve a finite limit as $\nc \ra \infty$, and also from the fact that the number of D7-branes must be of the same order as the number of D3-branes since they produce an ${O}(1)$-modification of the geometry. The power of the `t Hooft coupling is the same as that encountered in \cite{back,back2,back3} and is a strong-coupling result.

\subsection{Chemical potential}
\label{chemical}
In the `magnetic description' in terms of the axion one cannot introduce a chemical potential $\Phi$ conjugate to the number of D7-branes \cite{hawking-ross}. In other words, in this description one is restricted to work with a constant total number of D7-branes. In order to explore the complete thermodynamics of the system, we must first clarify how a non-constant number of D7-branes can be described by working with the field with respect to which the D7-branes are electrically charged. In ten-dimensional type IIB string theory this is the RR 8-form defined (roughly) through $dC_8 \sim \star d\chi$, where $\star$ denotes the Hodge dual in ten dimensions. It is easy to see from this relation that  $C_8$ may be taken to have only one non-zero component  
$C_{txy\theta^1\ldots\theta^5} (u)$, where $\theta^\alpha$ are coordinates on the $S^5$. This confirms our intuition that the D7-branes are wrapped on the $S^5$, extend along the $txy$-directions and are smeared along the $z$-direction.  Thus from the effective five-dimensonal viewpoint the D7-branes behave as a 2-brane charge density oriented in the $xy$-directions and distributed along the $z$-direction. 

Upon reduction on the $S^5$, $C_8$ gives rise to a 3-form that we will denote simply as $C$. Its precise definition in the effective five-dimensional theory follows from the requirement that its Bianchi identity be precisely the equation of motion for the axion resulting from the action \eqn{action} (and vice versa). Thus 
\be
dC = e^{2\phi} * d\chi = a\, \frac{e^{7\phi/4} \sqrt{\cb}}{u^3} \,
dt\wedge dx \wedge dy \wedge du \,,
\label{fieldstrength}
\ee
where in this and subsequent equations $*$ denotes the five-dimensional Hodge dual with respect to the metric \eqn{sol1}. Changing to the FG coordinate via \eqn{uOFv}, writing \mbox{$C=C_{txy}(v) \, dt\wedge dx\wedge dy$}, and using the near-boundary expansions \eqn{expansionFunctions}, we see that asymptotically 
\be
C_{txy} (v) = - \frac{a}{2 v^2} + \left( \cc- \frac{2}{3} a^3 \log v \right) +  {O}(v^2) \,.
\label{ctxy}
\ee
The integration constant  $\cc$ is  fixed by the regularity condition that $C_{txy}$ vanish at the horizon \cite{finitebaryon}. 
The most straightforward way to understand this requirement  is to continue to Euclidean signature. In that case $dt_\mt{E}$ is ill-defined at $u=u_\mt{H}$ (the `tip of the cigar') and hence $C_{t_\mt{E}xy}$ must vanish there. Equivalently, in Lorentzian signature $dt$ is ill-defined at the bifurcation surface of the Killing horizon, and therefore $C_{txy}$ must vanish at the horizon. 

We must now pause to appreciate an important difference between a 1-form and a 3-form gauge potential in an asymptotically AdS$_5$ spacetime. The time component of a 1-form potential falls off as 
\be
A_t = c_1 + c_2 v^2 + \cdots \,.
\ee
In this case, up to a normalization constant, one identifies the coefficient of the leading term, $c_1$, with the chemical potential, and the coefficient of the subleading term, $c_2$, with the charge density. 
This  identification is consistent with the fact that in this case $* dA$ near the boundary of $AdS_5$, whose integral measures the total charge, is proportional to $c_2$ and independent of $c_1$.  In contrast, in the case of a 3-form, one must identify the coefficient of the leading $1/v^2$ term (times $-2$) with the charge density, as is clear from the fact that this is proportional to $* dC \sim a$ and independent of $\cc$. Correspondingly, one must identify the subleading constant term $\cc$ with the chemical potential, up to a normalization constant. As we will see below, the normalization 
\be
\Phi = - \frac{\cc}{2 \kappa^2} = -\frac{\nc^2 \, \cc}{8\pi^2} 
\label{potential}
\ee
is the correct one, in the sense that the potential defined in this way agrees exactly for all values of $a$ and $T$ with the thermodynamic quantity defined in \eqn{sphi2}, provided we choose $\cs=-1$. Details can be found in the next section and in Appendix \ref{highApp}. Suffice it to note here that the fact that one must choose a particular value of $\cs$ is not at all surprising. Indeed, a renormalized `electric' action in which the axion had been  traded in favour of the 3-form $C$ would still suffer from scheme-dependence associated to the possible addition of finite counterterms. By writing eqn.~\eqn{potential} we have implicitly made a particular choice of scheme in this electric formulation, as is clear from the fact that the rescaling  \eqn{resc} would shift $\cc$ due to the logarithmic term in \eqn{ctxy}. Thus it is not surprising that the definitions \eqn{potential} and  \eqn{sphi2} coincide only provided a particular choice of scheme is also made for the free energy $F$ that enters the latter equation. We will verify below that, once this choice is made, any subsequent change of scheme respects this agreement.

\subsection{Fundamental thermodynamic identities}
\label{fund}
As usual, the on-shell classical action \eqn{on} provides a saddle-point approximation to the corresponding thermodynamic potential through the identification 
\be
S_\mt{on-shell} = \beta \int d^3 x \, F \,,
\ee
where $\beta=1/T$. In this case, since we are working at constant temperature and charge density, $F(T,a)$ is the free energy density in the canonical ensemble (see e.g.~\cite{hawking-ross}). Therefore $F$ is related to the energy density through 
\be
F=E-Ts \,,
\label{F}
\ee
and it obeys 
\be
dF = -s dT + \Phi da \,,
\label{dF}
\ee
which together with \eqn{F} is of course equivalent to the first law of thermodynamics
\be
dE= Tds + \Phi da \,.
\ee
Note that eqn.~\eqn{dF} defines the chemical potential as the intensive variable conjugate to the charge density $a$, namely we have
\bea
s&=& - \left( \frac{\partial F}{\partial T} \right)_a \,, \label{sphi1} \\
\Phi &=& \left( \frac{\partial F}{\partial a} \right)_T \,. 
\label{sphi2}
\eea
We will verify below that this definition of $\Phi$ agrees with the definition \eqn{potential}. Note that $a$ has dimensions of energy and that $\Phi$ has dimensions of energy/length$^2$. This is appropriate since $a$ measures the number of D7-branes per unit length in the $z$-direction, and $\Phi$ measures the energy cost per unit $xy$-area of introducing a D7-brane that extends in those directions --- see Appendix \ref{FPrelations}.

The thermodynamic potential in the grand canonical ensemble is simply given by
\be
G = E -Ts - \Phi a \,,
\label{G}
\ee
and it obeys
\be
dG = -s dT - a d\Phi \,.
\ee
Presumably, $G$ would equal the on-shell action computed in terms of the 3-form $C$ dual to $\chi$ \cite{hawking-ross}, once an appropriate renormalization procedure had been applied in such an `electric formulation'. 

Extensivity of the energy $E(s,a)$ as a function of its natural variables implies that the thermodynamic potentials are related to the pressures through (cf.~\cite{blackfold4})
\bea
F&=& -P_{xy} \,, \label{FF} \\
G&=&-P_z \,, \label{GG} 
\eea
which implies that the pressure difference is sustained by the presence of a non-zero chemical potential, namely 
\be
P_z - P_{xy} = \Phi a \,.
\label{deltap}
\ee
The relation $G=-P$ is familiar in the isotropic case. A derivation of the relations \eqn{FF}-\eqn{GG} in the anisotropic case is given in Appendix \ref{FPrelations}.

An important consistency check is the fact that all the thermodynamic identities above are scheme-independent, meaning that they are invariant under rescalings of $\mu$. Consider for example the relation~\eqn{F}. From eqns.~\eqn{on} and \eqn{counterterms} it is clear that under the rescaling \eqn{resc} $F$ transforms as
\be
F (k a, k T) = k^4 \, F (a,T) +  
k^4 \log k \, \frac{\nc^2 a^4}{48\pi^2}  \,,
\ee
which immediately implies that  $F$ must take the form
\be
F(a,T) = a^4 f \left( \frac{a}{T} \right)  
+ a^4 \log \left(\frac{a}{\mu} \right) \, \frac{\nc^2}{48\pi^2} \,.
\label{together}
\ee
Therefore 
\be
\frac{\partial F}{\partial \mu} = \frac{\partial E}{\partial \mu} \,,
\ee
consistently with the fact that $s$ is scheme-independent, 
\be
\frac{\partial s}{\partial \mu} = 0 \,,
\label{smu}
\ee
as expected since it is evaluated at the horizon. In contrast, we emphasize that  the chemical potential \emph{is} scheme-dependent. This is suggested by the fact that it cannot be obtained with horizon information alone (since in fact regularity implies that it must vanish there) but must be obtained from the fall-off \eqn{ctxy}. The scheme-dependence of $\Phi$ is also implied by its thermodynamic definition \eqn{sphi2}, which together with \eqn{together} implies that 
\be
\Phi (a,T) = a^3 \varphi \left( \frac{a}{T} \right)  
+ 4\, a^3 \log \left(\frac{a}{\mu} \right) \, \frac{\nc^2}{48\pi^2} \,.
\label{implies}
\ee
To see that this is consistent with the definition \eqn{potential}, we note from eqn.~\eqn{ctxy} that under the rescaling \eqn{resc} the constant term transforms as 
\be
\cc' = k^3 \cc - \frac{2}{3} k^3 \log k \,, 
\ee
where the $k^3$ factor comes from the transformation of $dt\wedge dx\wedge dy$. Consequently, the potential transforms as 
\be
\Phi (k a, k T)= k^3 \Phi (a,T) 
+ 4 k^3 \log k \, \frac{\nc^2}{48\pi^2} \,,
\ee
which implies \eqn{implies}. In summary, eqs.~\eqn{later}, \eqn{together} and \eqn{implies} imply that 
\be
\frac{\partial E}{\partial \mu} = - \frac{\partial P_{xy}}{\partial \mu} = 
\frac{1}{3} \frac{\partial P_z}{\partial \mu} = \frac{a}{4} \frac{\partial \Phi}{\partial \mu} \,,
\ee
which together with \eqn{smu} guarantee that the thermodynamic identities \eqn{F}, \eqn{G}, \eqn{FF} and \eqn{GG} are scheme-independent. 

Incidentally, we note that the relations above, together with the anomaly equation 
\be
\langle T^i_i \rangle = -E + 2P_{xy} + P_z = {\cal A} \,,
\ee
imply that 
\be
E - \frac{3}{4} T s - \frac{1}{4} \Phi a = \frac{1}{4} {\cal A} \,.
\label{violation}
\ee
As discussed e.g.~in \cite{comp5,comp6},\footnote{The factor of $1/4$ in front of $\Phi a$ differs from the $3/4$ factor in those references (in four spacetime dimensions) because here $a$ has dimensions of mass instead of mass$^3$.} if the only two scales were truly the temperature and the chemical potential then the first law would imply that the combination on the left-hand side would vanish identically. The fact that this is not the case here is another manifestation of the presence of an additional scale (the reference scale $\mu$), as implied by the anomaly.  

To close this section, we recall  that the necessary and sufficient conditions for \emph{local} thermodynamic stability in the canonical ensemble can be written as \cite{LL}
\bea
c_a &\equiv& \left(\frac{\partial E}{\partial T} \right)_a 
= T \left( \frac{\partial S}{\partial T} \right)_a  > 0 \,, \label{ca} \\
\Phi' &\equiv& \left(\frac{\partial \Phi}{\partial a} \right)_T = 
\left(\frac{\partial^2 F}{\partial a^2} \right)_T > 0 \,. \label{phia}
\eea
The first condition is just the statement that the specific heat at constant charge is positive. The second one is the statement that the system is stable against infinitesimal charge fluctuations.  

\section{Phase diagram}
\label{sec-phasediagram}

\begin{figure}
\begin{center}
\includegraphics[width=0.99\textwidth]{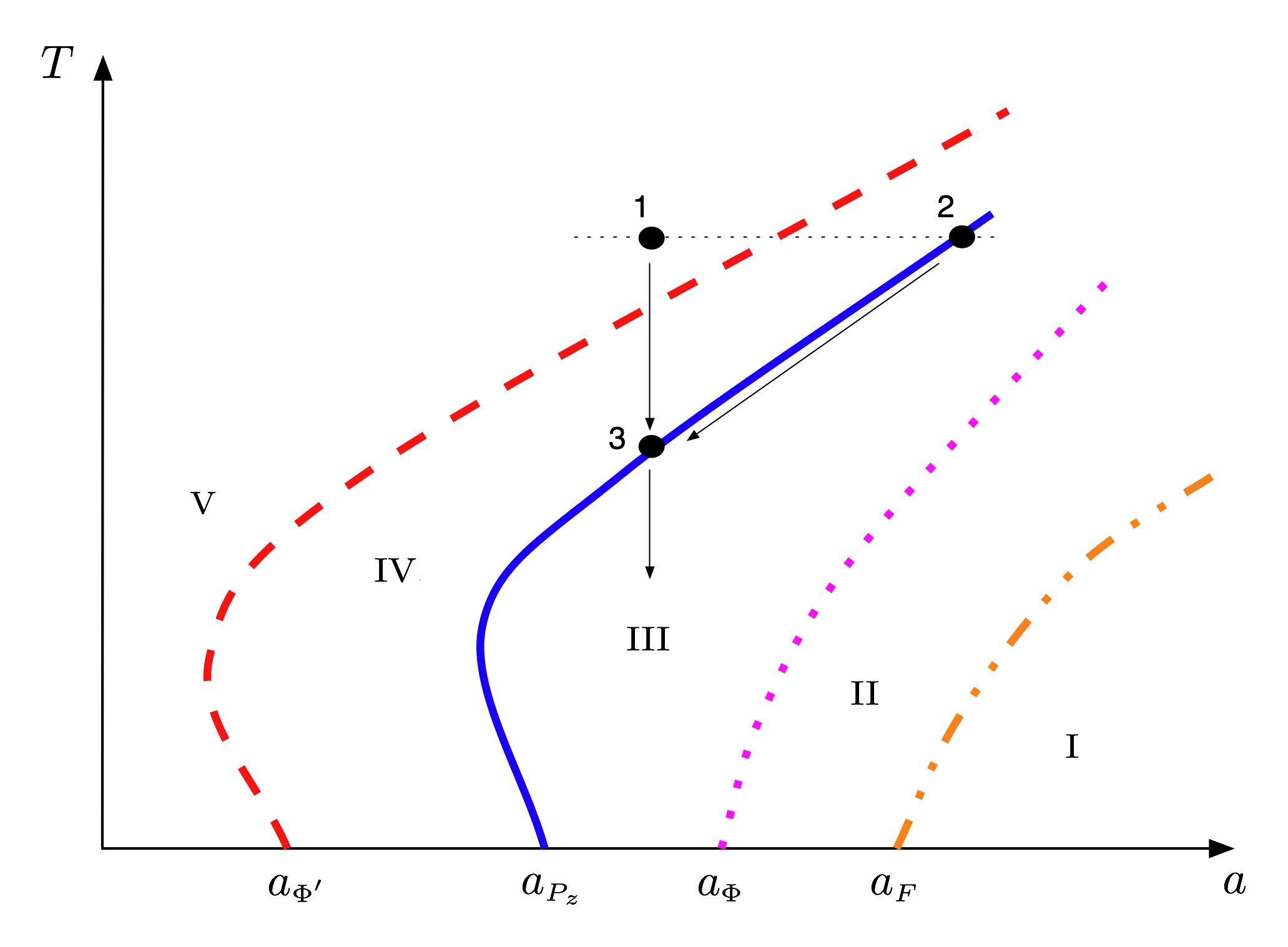}
\caption{\small Qualitative phase diagram. The orange, dashed-dotted curve is defined by the condition $F(a,T)=F^0(a,T)$. The magenta, dotted curve is defined by the condition $\Phi(a,T)=0$. The blue, continuous curve is defined by the condition $P_z (a,T) = P^0(T)$. The dashed, red curve is defined by the condition $\Phi' (a,T) = 0$. In Zones I, II and III the system is in a homogeneous phase, whereas it is in a mixed phase in Zones IV and V.  See the main text for more details.}
\label{phasediagram}
\end{center}
\end{figure}

Above we reviewed and clarified certain aspects  which we will now apply to the thermodynamics of our specific system. In order to do so we will examine the limits of high and low temperature. This information alone will suffice to construct the qualitative phase diagram shown in Fig.~\ref{phasediagram}.

In the high-temperature limit, $T\gg a, \mu$, one can obtain approximate analytical expressions for the various quantities of interest by perturbing around the isotropic solution \eqn{isotropic}. The details can be found in Appendix \ref{highApp}. Here we simply collect the results for the energy density, the pressures and the entropy density. Omitting terms of $ {O}(a^6)$ these are
\bea
E &=& E^0 (T) + \frac{\nc^2 T^2 a^2}{32}  + 
\frac{\nc^2 a^4}{1536 \pi^2} \left[ 8 \cs -41 - 32 \log \left( \frac{a}{2\pi T} \right) \right]
+ \frac{\nc^2 a^4}{48 \pi^2} \log \left( \frac{a}{\mu} \right) \,,\,\,\,\, \label{Ehigh}\\
P_{xy} &=& P^0 (T)  +  \frac{\nc^2 T^2 a^2}{32}  + 
\frac{\nc^2 a^4}{1536 \pi^2} \left[ -8 \cs +9 + 32 \log \left( \frac{a}{2\pi T} \right) \right]
- \frac{\nc^2 a^4}{48 \pi^2} \log \left( \frac{a}{\mu} \right) \,,\,\,\,\, \label{Pxyhigh} \\
P_z &=& P^0 (T)  -  \frac{\nc^2 T^2 a^2}{32}  + 
\frac{\nc^2 a^4}{512 \pi^2} \left[ 8 \cs -9 - 32 \log \left( \frac{a}{2\pi T} \right) \right]
+ 3 \,\frac{\nc^2 a^4}{48 \pi^2} \log \left( \frac{a}{\mu} \right) \,,\,\,\,\, \label{Pzhigh} \\
s &=& s^0 (T)  +  \frac{\nc^2 T a^2}{16}  - 
\frac{\nc^2 a^4}{48 \pi^2 T}  \,,
\eea
where $E^0 (T), P^0 (T)$ and $s^0 (T)$ are the corresponding results for the undeformed (i.e.~$a=0$) ${\cal N}=4$ SYM theory given in \eqn{entropyN=4} and \eqn{EP0}. The sign of the leading, $ {O}(a^2)$ correction to these results is consistent with the positive slope of the $E$ and $P_{xy}$ curves and the negative slope of the $P_z$ curve in Fig.~\ref{enpress_aoverT} at small $a/T$. The chemical potential can be obtained through 
eqn.~\eqn{sphi2} and it reads 
\be
\Phi =  -  \frac{\nc^2 T^2 a}{16}  + 
\frac{\nc^2 a^3}{384 \pi^2} \left[ 8 \cs -9 - 32 \log \left( \frac{a}{2\pi T} \right) \right]
+ 4 \, \frac{\nc^2 a^3}{48 \pi^2} \log \left( \frac{a}{\mu} \right) \,.
\label{PhiHigh}
\ee

In the opposite limit, $T\ll a,\mu$, one can also obtain approximate analytical expressions for the various quantities of interest. The details can be found in Appendix \ref{lowApp}. Here we simply collect the results for the energy density, the pressures and the entropy density:
\bea
E &=& 
\frac{\nc^2 a^4}{5376 \pi^2} \left( 28 \cs -1 +192 c_\mt{int} \right) 
+ \frac{\nc^2 a^4}{48 \pi^2} \log \left( \frac{a}{\mu} \right) 
+ \frac{8 c_\mt{ent}}{11} \nc^2 a^{1/3} T^{11/3} + \cdots \,, \label{E_low} \\
P_{xy} &=& 
- \frac{\nc^2 a^4}{5376 \pi^2} \left( 28 \cs -1 +192 c_\mt{int} \right) 
- \frac{\nc^2 a^4}{48 \pi^2} \log \left( \frac{a}{\mu} \right) 
+ \frac{3  c_\mt{ent}}{11} \nc^2 a^{1/3} T^{11/3} + \cdots \,, \label{pxylow}\\
P_z &=&
\frac{\nc^2 a^4}{5376 \pi^2} \left( 84 \cs +109 +576 c_\mt{int} \right) 
+3\, \frac{\nc^2 a^4}{48 \pi^2} \log \left( \frac{a}{\mu} \right) 
+ \frac{2c_\mt{ent}}{11} \nc^2 a^{1/3} T^{11/3} + \cdots \,, \,\,\,\,\,\,\,\,
\label{pzlow} \\
s &=& c_\mt{ent} \nc^2 a^{1/3} T^{8/3} + \cdots \,,
\eea
where we have omitted terms of higher order in an expansion in $T/a$. In these expressions $c_\mt{int}$ is an integration constant that could be determined numerically, and $c_\mt{ent}>0$ is the constant defined in \eqn{slif}. The specific values of these constants will have no effect on the qualitative physics. The chemical potential is 
\be
\Phi=4\, \frac{\nc^2 a^3}{5376 \pi^2} \left( 28 \cs +27 +192 c_\mt{int} \right) 
+4\, \frac{\nc^2 a^4}{48 \pi^2} \log \left( \frac{a}{\mu} \right) 
- \frac{c_\mt{ent}}{11} \nc^2 a^{-2/3} T^{11/3} + \cdots \,.
\label{philow}
\ee
Note that at $T=0$ one has\footnote{In this limit the horizon disappears and the string-frame metric develops a naked curvature singularity deep in the IR (see Section \ref{discussion} and Appendix \ref{comparison} for details). To avoid this one can simply imagine that $T$ is made very small but not zero. \label{foot}}
\be
s=0 \sac E = -P_{xy} \sac 3 P_{xy} + P_z = \frac{\nc^2 a^4}{48 \pi^2} \,.
\label{zero}
\ee
These results are in agreement with Fig.~\ref{enpress_Tovera}, where in particular we see that $E$ and $P_{xy}$ have opposite signs in the limit $T/a \ra 0$. The signs of $E, P_{xy}$ and $P_z$ at low $T/a$ depend on the value of the ratio $a/\mu$. For example, the energy becomes negative as $T\ra 0$ provided $a$ is sufficiently smaller than $\mu$. We emphasize that this is not an inconsistency: It means that the energy can be lower than that of the undeformed ${\cal N}=4$ theory at zero temperature, but the energy is still bounded from below for all values of $a$, and achieves this absolute minimum at a scale set by $\mu$. We will come back shortly to the meaning of the signs of the pressures.

The coefficients of the $\log(a/\mu)$ terms above, both at high and at low temperature, are in perfect agreement with those in eqs.~\eqn{later} and \eqn{implies}, as is the trace of the stress tensor
\be
\langle T_i^i \rangle = -E + 2P_{xy} + P_z =\frac{\nc^2 a^4}{48 \pi^2} 
\ee
with eqs.~\eqn{trace}-\eqn{anomaly}. Also, with the explicit expressions above in hand, it is straightforward to verify that all the thermodynamic identities of Section \ref{fund} are identically satisfied in both limits.

\begin{figure}[h]
\begin{center}
\includegraphics[width=0.8\textwidth]{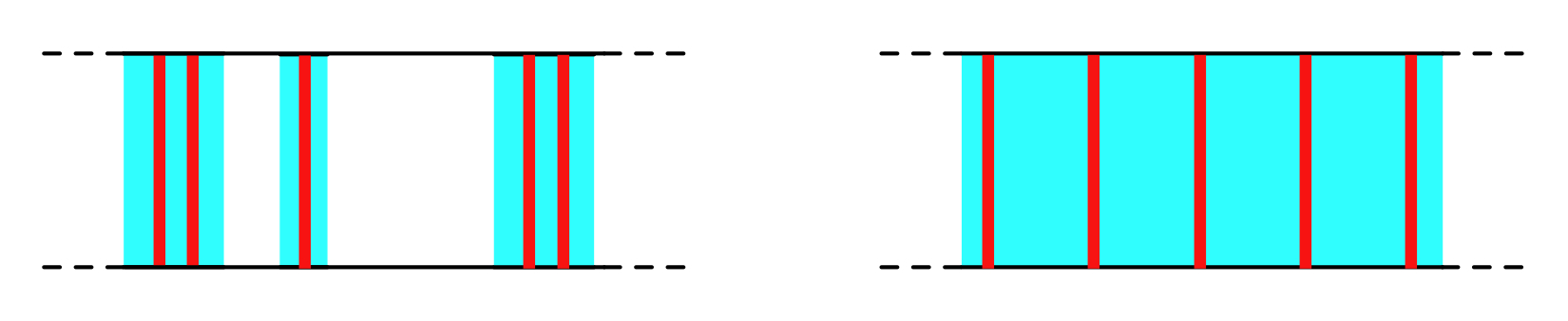}
\caption{\small (Left) Inhomogeneous mixed phase in which isotropic and anisotropic regions coexist. (Right) Homogeneous anisotropic phase.}
\label{redistribution2}
\end{center}
\end{figure}

We are now ready to understand the phase diagram qualitatively. We begin by examining the system at zero temperature and finite density. In this case we see from eqs.~\eqn{pxylow}, \eqn{pzlow} and \eqn{philow} that
\bea
3F &=& 
 \frac{\nc^2 a^4}{5376 \pi^2} \left( 84 \cs - \,\,\,\,\,\,3 +576 c_\mt{int} \right) 
+ 3\, \frac{\nc^2 a^4}{48 \pi^2} \log \left( \frac{a}{\mu} \right) \,,\label{pxy0} \\
\frac{3}{4} \Phi&=& \frac{\nc^2 a^3}{5376 \pi^2} \left( 84 \cs +\,\, 81 +576 c_\mt{int} \right) 
+ 3 \, \frac{\nc^2 a^4}{48 \pi^2} \log \left( \frac{a}{\mu} \right) \,, \label{phi0}\\
P_z &=&
\frac{\nc^2 a^4}{5376 \pi^2} \left( 84 \cs +109 +576 c_\mt{int} \right) 
+3\, \frac{\nc^2 a^4}{48 \pi^2} \log \left( \frac{a}{\mu} \right) \,, \label{pz0} \\
\frac{1}{4} \Phi' &=& \frac{\nc^2 a^2}{5376 \pi^2} \left( 84 \cs +193 +576 c_\mt{int} \right) 
+  3\, \frac{\nc^2 a^2}{48 \pi^2} \log \left( \frac{a}{\mu} \right)  \,,
\label{dphi}
\eea
where we recall that $\Phi' = \partial \Phi/\partial a$ and $F=-P_{xy}$. Simple inspection of these equations shows that $F, \Phi, P_z$ and $\Phi'$ are all positive for $a\gg\mu$, but that as $a$ gradually decreases they become negative sequentially in the order in which they are listed above, at values $a_\mt{$F$} > a_\mt{$\Phi$} > a_\mt{$P_z$} > a_\mt{$\Phi'$}$. Note that the precise values of $a_i$ depend on $\cs$ and $c_\mt{int}$, but that their ordering is independent of these constants, and that the scale of all the $a_i$ is set by $\mu$. It follows that at zero temperature and sufficiently low densities, $a<a_\mt{$\Phi'$}$, the condition \eqn{phia} is violated and therefore the system is unstable against infinitesimal  charge fluctuations, i.e.~small clumps of charge with density slightly higher than the average will grow instead of relaxing back to the average. At densities 
$a_\mt{$\Phi'$} < a < a_\mt{$P_z$}$ the system becomes stable against infinitesimal charge fluctuations, and therefore we will refer to this phase as metastable. Yet, the system is still unstable against finite charge fluctuations, as signaled by the fact that if $a < a_\mt{$P_z$}$ then the pressure of the isotropic phase is higher than that of the anisotropic phase:\footnote{In this  and in subsequent equations we will not set $T=0$ explicitly because, as we will see, this instability extends to $T>0$.  In order to specify to $T=0$ one must simply remember that $P^0(T)=0$.} 
\be
P_z(a,T) < P^0(T) \,. 
\label{pcondition}
\ee
As a consequence, bubbles of the isotropic phase can form and grow inside the anisotropic phase, forcing a redistribution of the total charge into a smaller volume.\footnote{Note that no charge redistribution can occur in the $xy$-directions, since the branes are extended along these directions.} In other words, even if one could somehow prepare a homogenous phase of density $a < a_\mt{$P_z$}$, this would fall apart into a mixed phase consisting of `droplets' or `filaments' (or, more precisely, thick membranes) of anisotropic phase with density $a = a_\mt{$P_z$}$ and $P_z=0$  surrounded by vacuum regions with $a=P^0=0$, as depicted in Fig.~\ref{redistribution2}(left). 
Note that the chemical potential satisfies $\Phi(a_\mt{$P_z$})<0$.  This is an important consistency condition, for it means that it is indeed advantageous for the charge to stay together in the anisotropic region; we will comment further on this below.  At densities $a > a_\mt{$P_z$}$ the preferred phase is a homogeneous anisotropic phase, as shown in Fig.~\ref{redistribution2}(right), which is stable. This physics is similar to that of zero-temperature QCD at finite baryon density \cite{colorsuperconductivity,finite2,finite3}, which suggests that the phase transition from the mixed phase to the homogenous anisotropic phase may occur by a percolation mechanism as the anisotropic regions merge, again in analogy to models of restoration of chiral symmetry in QCD \cite{percolation,percolation2}. We will elaborate on this similarities in Section  \ref{discussion}. 
 
The dynamics of the isotropic bubbles may be complicated and will depend, among other things, on their surface tension.   We will come back to this point in Section \ref{discussion}. Here we just note that the `mechanical condition' \eqn{pcondition} is exactly equivalent to the 
condition that a compression of the  charge is thermodynamically preferred. Indeed, let $L_z$ be the length of the $z$-direction, which can be taken to infinity at the end.\footnote{Note that the $\nc \ra \infty$ limit in which we are working ensures that we are in the thermodynamic limit even if the total volume $V=L_x L_y L_z$ is finite.} We wish to fix the total D7-brane charge $\bar a$ (we follow the notation of Appendix \ref{FPrelations}) and compare the free energies of the following two configurations (see Fig.~\ref{redistribution}): 
\begin{figure}[h!]
\begin{center}
\includegraphics[width=0.7\textwidth]{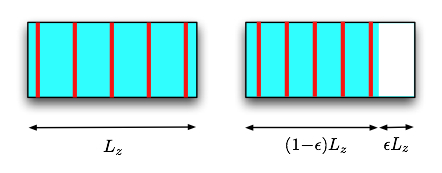}
\caption{\small Redistribution of D7-brane charge along the $z$-direction.}
\label{redistribution}
\end{center}
\end{figure}
(i) a configuration in which the charge is distributed homogeneously over the entire length $L_z$ with density $a=\bar a/L_z$, and (ii) a configuration in which a small fraction $\epsilon L_z$ of the total length is occupied by the isotropic phase and the rest is occupied by the anisotropic phase with the total charge distributed homogeneously but with a higher density 
\be
a'=\frac{\bar a}{(1-\epsilon)L_z} \simeq a + a \epsilon \,.
\ee 
The free energy (per unit $xy$-area) of each configuration is 
\bea
F_{(i)}&=& L_z F(a,T)\nonumber \\
F_{(ii)} &=&  L_z(1-\epsilon)\, F(a+ a \epsilon,T) + \epsilon L_z\, F^0(T) 
\nonumber \\
&\simeq &   L_z F(a,T)  + \epsilon L_z F^0(T) 
-\epsilon L_z \left( F(a,T) - \frac{dF(a,T)}{da} a \right) \nonumber \\
&= &   L_z F(a,T)  - \epsilon L_z P^0(T)  
+ \epsilon L_z \Big( P_{xy}(a,T) + \Phi (a,t) a \Big) \nonumber \\
&= &  L_z F(a,T)  - \epsilon L_z P^0(T) + \epsilon L_z P_z (a,T) \,,
\eea
where we have used eqs.~\eqn{sphi2}, \eqn{FF} and \eqn{deltap}. We therefore conclude that $F_{(ii)} < F_{(i)}$ if and only if \eqn{pcondition} is satisfied.

In order to understand how the physics above extends to or is modified at $T>0$, let us examine the limit $T \gg a$. Before addressing the behaviour under charge fluctuations, we briefly establish that there are no thermal instabilities. Indeed, ignoring terms of $ {O}(a^4)$ in \eqn{Ehigh} we see that the specific heat \eqn{ca} is positive at high temperatures. 
The first temperature-dependent correction at low temperatures in eqn.~\eqn{E_low} also yields $c_a>0$. We see from the positive slope of the red, continuous curve in Fig.~\ref{enpress_Tovera} that this extends to all values of $T/a$. In fact, not only the energy but  also the pressures are monotonically increasing functions of $T$ for fixed $a$. This implies that the speed of sound (squared) in the $xy$- and $z$-directions is positive: 
\be
v_{xy}^2 = \left( \frac{\partial P_{xy}}{\partial E} \right)_a > 0 \sac 
v_{z}^2 = \left( \frac{\partial P_{z}}{\partial E}\right)_a > 0\,.
\label{speed}
\ee 
Returning to charge fluctuations, we see from \eqn{PhiHigh} that 
\be
\Phi' \equiv \left( \frac{\partial \Phi}{\partial a}\right)_T \simeq 
 -  \frac{\nc^2 T^2}{16}  < 0
\ee
at sufficiently high temperatures, demonstrating that the local stability condition \eqn{phia} is violated in this limit. We also note from eqn.~\eqn{Pzhigh} that the leading $ {O}(a^2)$ correction to $P_z$ is negative, meaning that we are again in the situation described by eqn.~\eqn{pcondition}.

With this information we can now draw the qualitative phase diagram --- see Fig.~\ref{phasediagram}. 
There are four curves defined (from right to left) by the conditions $F(a,T)-F^0(a,T)=0$ (orange, dashed-dotted curve), $\Phi(a,T)=0$ (magenta, dotted curve), \mbox{$P_z(a,T)-P^0(T)=0$} (blue, solid curve) and $\Phi' (a,T)=0$ (red, dashed curve). These four curves intercept the horizontal, zero-temperature axis at the points $a_i$ defined above. The slopes at $T=0$ of the first two curves are positive, whereas the slopes of the other two are negative. This can be seen by examining the leading temperature correction to the zero-temperature results \eqn{pz0}-\eqn{dphi}. Consider for example the chemical potential. Let $(n_T,n_a)$ be the tangent vector to the curve $\Phi(T,a)=0$ at the point $(T,a)=(0,a_\mt{$\Phi$})$. By definition the derivative of $\Phi$ along the curve vanishes, namely
\be
0 = n_T \frac{\partial \Phi}{\partial T} + n_a \frac{\partial \Phi}{\partial a} \,.
\ee 
Obviously at $(T,a)=(0, a_\mt{$\Phi$})$ we have $n_T>0$, and we see from eqn.~\eqn{philow} that $\partial \Phi/\partial T <0$. Also, we know that $\partial \Phi/\partial a >0$ since $\Phi$ goes from negative to positive  across the curve in the direction of increasing $a$. It follows that we must have $n_a >0$. A similar argument applies to the other three curves. 

Note that the four curves must `extend to infinity' in the sense that they must contain points for which both $T$ and $a$ are arbitrarily large. To see this, consider again the chemical potential. This is negative in the limit $T\gg a, \mu$ but positive in the limit $a\gg T,\mu$. Thus if we start at any point $(a,T)$ at which $\Phi$ is negative and we increase $a$ keeping $T$ fixed, eventually we reach the regime in which $a\gg T,\mu$ where $\Phi$ must be positive. Thus a horizontal, constant-$T$ line always crosses the curve $\Phi=0$ no matter how large $T$ is. Similarly, if we start at a point $(a,T)$ where $\Phi$ is positive and we increase $T$ keeping $a$ fixed, eventually we reach the regime in which $T\gg a,\mu$ where $\Phi$ must be negative. Thus a vertical, constant-$a$ line always crosses the curve $\Phi=0$ no matter how large $a$ is (specifically, provided it is larger than $a_\mt{$\Phi$}$). An analogous argument applies to each of the other three curves. Of course, from the asymptotic information at large and small $a/T$ we cannot deduce the detailed form of the curves at arbitrary values of $a/T$;  for example, we cannot exclude that some curves cross each other. However, we have verified the qualitative form of the phase diagram shown in Fig.~\ref{phasediagram} with our numerical solution. In particular, we have verified that the curves do not cross.

Returning to the phase diagram, we conclude that there are five distinct zones:
\bea
\mbox{I:}&& 
F(a,T) > F^0(T) \,, \;\;\; \Phi(a,T) > 0 \,, \;\;\; P_z (a,T) > P^0(T) 
\,, \;\;\; \Phi' (a,T) > 0 
\,, \nonumber \\
\mbox{II:}&& 
F(a,T) < F^0(T) \,, \;\;\; \Phi(a,T) > 0 \,, \;\;\; P_z (a,T) > P^0(T) 
\,, \;\;\; \Phi' (a,T) > 0 
\,, \nonumber \\
\mbox{III:}&& 
F(a,T) < F^0(T) \,, \;\;\; \Phi(a,T) < 0 \,, \;\;\; P_z (a,T) > P^0(T) 
\,, \;\;\; \Phi' (a,T) > 0 
\,, \nonumber \\
\mbox{IV:}&&  
F(a,T) < F^0(T) \,, \;\;\; \Phi(a,T) < 0 \,, \;\;\; P_z (a,T) < P^0(T) 
\,, \;\;\; \Phi' (a,T) > 0 
\,, \nonumber \\
\mbox{V:}&& 
F(a,T) < F^0(T) \,, \;\;\; \Phi(a,T) < 0 \,, \;\;\; P_z (a,T) < P^0(T) 
\,, \;\;\; \Phi' (a,T) < 0  \,. \;\;\; 
\label{region}
\eea
In Zones I, II and III the thermodynamically preferred configuration is a stable homogeneous phase. In contrast, the homogeneous phase is metastable in Zone IV and unstable in Zone V. In these zones the preferred configuration is a mixed, inhomogeneous phase in which multiple regions of the isotropic and the anisotropic phases coexist with each other, as in Fig.~\ref{redistribution2}. The local charge density in each of the anisotropic regions is $a_\mt{ani}>a$ and the pressure exactly equals that of the isotropic phase at the same temperature, 
$P_z(a_\mt{ani},T)=P^0(T)$. The pair $(a_\mt{ani},T)$ therefore lies on the blue, continuous curve of the phase diagram. Note that the chemical potential need not be, and in fact is not, the same in the isotropic and anisotropic regions: $\Phi$ vanishes in the isotropic phase and is negative in the anisotropic phase for points that lie on the blue curve. As mentioned above, this is an important consistency condition for the coexistence of the two phases. In a system in contact with an infinite  charge reservoir the chemical potential will eventually equilibrate to the same value everywhere. However, in the system under consideration, once an anisotropic region is surrounded by isotropic charge-less regions, the anisotropic region cannot draw additional charge from the latter. In this case consistency only requires that the chemical potentials verify $\Phi_\mt{ani} \leq \Phi_\mt{iso}$, since otherwise the charge would escape from the anisotropic to the isotropic regions. 

If we vary the temperature of a mixed-phase configuration adiabatically, $P^0(T)$ will change smoothly and the volume occupied by the anisotropic regions will adjust so that the pair $(a_\mt{ani},T)$ stays on the blue curve. For example, if we start with a density and temperature $(a,T)$ corresponding to the point `1' in the phase diagram, the corresponding pair $(a_\mt{ani},T)$ describing the anisotropic regions will be point `2'. As we lower the temperature, the pair $(a,T)$ moves down along a vertical line, whereas $(a_\mt{ani},T)$ moves down and leftwards along the blue curve. The two trajectories meet at point `3'. At this point the two densities agree, $a=a_\mt{ani}$, and thus the anisotropic phase fills out the entire space, i.e.~we have reached the homogeneous phase. As we keep lowering the temperature we stay in the homogeneous phase. Because of the form of the blue curve, it is clear that if we start with a density $a$ to the left of the turning point of the blue curve then as we vary the temperature we never reach the homogeneous phase. Similarly, for an intermediate range of initial densities between the turning point and  $a_\mt{$P_z$}$, we may enter and subsequently leave the homogeneous phase.


\section{Discussion}
\label{discussion}
In this paper we have studied the finite-temperature generalization of the type IIB supergravity solution of \cite{ALT} dual to an anisotropic deformation of ${\cal N}=4$ SYM. The solution obeys  $AdS_5 \times S^5$ boundary conditions, as appropriate for a marginally relevant deformation, possesses a regular anisotropic horizon, and is completely smooth on and outside the horizon. These features guarantee that it is solidly embedded in string theory. The deformation of the gauge theory is induced by a position-dependent $\theta$-parameter (see eqn.~\eqn{def}), which in the dual gravity description can be understood as resulting from a density of dissolved  D7-branes. 

The fact that the D7-branes are completely wrapped on the $S^5$ leads to a great technical simplification, because it implies that the $SO(6)$ symmetry of the undeformed theory is preserved. This means in turn that none of the Kaluza-Klein (KK) modes on the $S^5$ are excited, and therefore that the solution can be found by working directly with five-dimensional supergravity with a negative cosmological constant coupled to a few matter fields --- just the axion and the dilaton. 

At zero temperature the IR limit of the solution is given by eqs.~\eqn {metric_ALT}-\eqn{formal}. As explained in Appendix \ref{comparison} the string-frame metric possesses a naked singularity deep in the IR, i.e.~at $r\to 0$ in the coordinates of \eqn {metric_ALT} or at $u\to \infty$ in terms of the coordinate used above. The metric in the Einstein frame is also pathological. As pointed out in \cite{tidal} and elaborated upon in \cite{tidal2}, although all curvature invariants of the Einstein-frame metric are finite, tidal forces diverge as one approaches the $r= 0$ hypersurface. 
In contrast, at finite temperature the problematic region is hidden behind a horizon. It is remarkable that in this case the solution takes into account the full back-reaction of the D7-branes but exhibits no pathologies regardless of their number (density), which scales as  $n_\mt{D7} \sim \nc/\lambda$ (see eqn.~\eqn{thetan}) and can be larger than 24. 
This provides a completely explicit counterexample (at least in the presence of a negative cosmological constant) to the somewhat extended belief that 24 is an upper bound on the possible number of D7-branes. In particular, the dilaton is finite everywhere. In this respect it is interesting to note that in the IR Lifshitz geometry \eqn{metric_ALT}-\eqn{formal} the dilaton grows without bound towards the UV ($r\to \infty$), but this growth is tamed by UV-completing the Lifshitz geometry with an asymptotic AdS region. We note that the parameter controlling the backreaction of the D7-branes is not just the ratio of their number (density) to the number of D3-branes, but this ratio enhanced by a power of the `t Hooft coupling, $\lambda n_\mt{D7}/\nc$, as in the case \cite{back,back2,back3} of flavour D7-branes dual to quark degrees of freedom in the ${\cal N}=4$ SYM plasma . In the latter case, however, inclusion of the D7-branes' backreaction does lead to a singularity in the UV part of the geometry \cite{sing2,back3}. The dual gauge theory interpretation is that the addition of matter degrees of freedom to a conformal field theory makes the $\beta$-function positive and thus leads to a Landau pole in the UV. In contrast, in the present case we are \emph{not} adding new degrees of freedom because the D7-branes do not extend to the AdS boundary, so from this viewpoint it is perhaps not  surprising that the gravity solution does not exhibit any pathologies in the UV. 

It would be interesting to generalize our construction to dissolved branes of other dimensions. One could certainly consider other branes partially  wrapped on the $S^5$ in such a way that their non-compact directions coincide with some of, but not all, the gauge theory directions, since we want them to break isotropy. However, this may be technically challenging if the branes break too much of the $SO(6)$ symmetry of the sphere, since in this case the KK modes will become excited. In this respect it might be useful to smear the branes on the $S^5$ as to preserve the maximum amount of symmetry possible, along the lines of \cite{smear,smear2} (see \cite{smear3} for a review). An additional desirable requirement is that the branes responsible for the anisotropy do not change the UV properties of the theory, since this would destroy the AdS asymptotics, thus obscuring the microscopic nature of the dual theory (if it exists at all). Presumably, this  requirement will exclude lower dimensional branes, as illustrated by the dissolved D1-branes and strings of Refs.~\cite{nc,nc2}, which are dual to non-commutative deformations of the ${\cal N}=4$ SYM theory that do change the UV physics. It would also be interesting to generalize our construction to gauge theories in different numbers of dimensions. A step in this direction was already given in \cite{ALT}, where an anisotropic Lifshitz-like geometry based on the D4-D6 system was constructed.

Our main focus in this paper was the study of the thermodynamics of the system. For this purpose we obtained the holographic stress tensor particularizing the results of \cite{odin,Yiannis}, and showed that its transformation under the rescaling $(T,a) \ra (kT,ka)$ contains an  inhomogeneous piece (see eqn.~\eqn{inhom}) induced by the presence of a non-zero conformal anomaly ${\cal A}=\nc^2 a^4/48 \pi^2$ (see eqn.~\eqn{anomaly}). This implies that the quantum theory depends on an additional scale $\mu$, a remnant of the renormalization process, analogous to the subtraction point in QCD. The main result of our study is the phase diagram of the system in the density-temperature plane, shown in Fig.~\ref{phasediagram}. As noted above, in the strict $T=0$ limit the horizon disappears and the string-frame metric exhibits a naked curvature singularity deep in the IR.  In this region we expect stringy $\alpha'$-corrections to become important and therefore some of the details of the phase diagram in a thin horizontal slice that includes the $T=0$ axis may be modified. 

We identified zones of the phase diagram in which the homogeneous phase is metastable or unstable. All the instabilities are related to the tendency of the D7-brane charge to clump together. This follows from the fact that if the charge density in a given region is too small, then the pressure in the $z$-direction in that region becomes lower than the pressure of the isotropic phase at the same temperature. Once this happens, `bubbles' of the isotropic phase can form and grow, thus forcing a compression of the charge density into smaller regions, as depicted in Fig.~\ref{redistribution2}. We verified that this `mechanical conclusion' was precisely equivalent to the condition that the free energy of the system be minimized. Note that in our analysis we assumed that the total number of D7-branes was fixed, but that their local number density was not. In other words, we considered inhomogeneous phases, which is of course natural from the viewpoint of the thermodynamics of the gauge theory. In the context of AdS/CFT one could choose to fix the D7-brane density at each point, since this is specified by the non-normalizable mode of the axion. However, one might suspect that the instability relies on IR near-horizon physics. If this is the case, then it is possible that even 
if the axion non-normalizable mode $\chi^{(0)}$ is held fixed an instability may still arise through the tendency of the instanton density 
${\cal O}_\chi \sim \mbox{Tr} \, F\tilde F$ to develop an expectation value.

\begin{figure}[t]
\begin{center}
\includegraphics[scale=0.8]{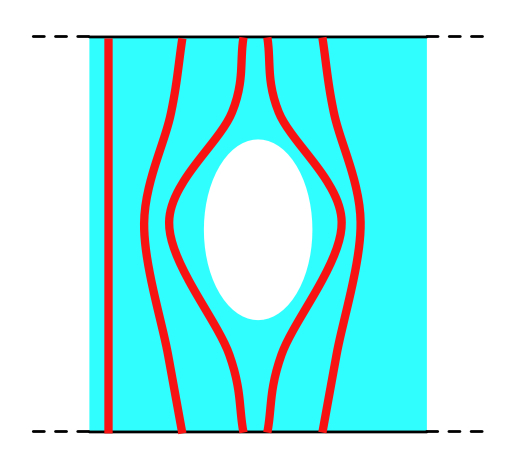}
\caption{\small Cartoon of the formation of a bubble of isotropic phase.}
\label{bubble}
\end{center}
\end{figure}

Analyzing the formation and dynamics of the isotropic bubbles would require consideration of many effects, in particular the bubble surface tension. This is certainly beyond the scope of this paper, so we will just  make two observations. First, we note that the isotropic bubbles can form even if the $xy$-directions are non-compact. A cartoon of how a typical bubble may start is shown in Fig.~\ref{bubble}.
The key point is that the bubble can form with a finite size in all directions, thus requiring a finite fluctuation, and subsequently expand, in particular in the $xy$-directions. 
Second, there should exist domain-wall solutions in the gravitational description which interpolate along the $z$-direction between regions with  $d \chi \neq 0$ and $d \chi = 0$, provided the pair $(T,a)$ in the former region lies on the blue curve of the phase diagram. These solutions would be dual to the inhomogeneous configurations depicted in Fig.~\ref{redistribution2}, and they would allow for an estimation of the surface tension of an anisotropic bubble. 
Presumably the instanton density $\langle {\cal O}_\chi \rangle \sim \langle \mbox{Tr} \, F\tilde F \rangle$
would be non-zero in the interface region.

An interesting extension on the gravity side would be to consider time-dependent solutions. The focus of this paper has been the static anisotropic solution, but one could now view this solution as an initial condition for a time-dependent solution. Specifically, one could look for a gravity solution such that the non-normalizable mode of the axion satisfies $d\chi\neq 0$ for $t< t_0$ and $d\chi =0$ for $t>t_0$. Presumably this would  allow one, for example, to compare the isotropization and the thermalization times for the resulting solution at $t>t_0$.

The physics displayed by our system in Zones IV and V is similar to that encountered in QCD at low temperature and finite baryon density \cite{colorsuperconductivity,finite2,finite3}. In that case one finds that the pressure of a chirally broken homogeneous phase with density lower than a critical density $n_0$ is negative.\footnote{Except in a tiny region of very small densities.} This indicates an instability of the system towards the formation of `droplets' of higher density $n_0$ (interpreted as nucleons) in which $P=0$ and chiral symmetry is restored, surrounded  by empty space with $n=0$ and $P=0$. In our case, the role of the chirally restored phase is played by the anisotropic phase, the analogue of $n_0$ is $a_\mt{$P_z$}$, and the `droplets' correspond to the `filaments', or more precisely thick membrane-like regions, of non-zero D7-brane density. These similarities suggest that the transition from the mixed phase to the homogeneous phase may occur via a percolation mechanism as the anisotropic regions merge, in analogy with some models of restoration of chiral symmetry in QCD \cite{percolation,percolation2}. Note however that the objects that percolate in our case have infinite extent in the $xy$-directions. In this respect a closer analogy might exist with suggested models of deconfinement based on percolation of extended colour strings \cite{strings,strings2}. 
The similarities between our system and QCD extend to sufficiently low temperatures, but not beyond: In our case the zones where the mixed phase exists (Zones IV and V) contain points with arbitrarily large temperatures and charge densities, whereas in QCD the mixed phase exists only in a low-temperature, low-density region of the phase diagram

The instabilities that we have uncovered are reminiscent of instabilities of weakly coupled plasmas in the presence of anisotropies, and we will now comment on some superficial similarities and differences between both cases. However, we emphasize from the start that addressing to what extent, if any, the instabilities encountered here can be viewed as a strong-coupling counterpart of weak-coupling instabilities is beyond the scope of this paper, and we make no claim to that effect in either direction. The ensuing comments should thus be taken merely as suggestive observations. 

\begin{figure}[t]
\begin{center}
\includegraphics[width=0.8\textwidth]{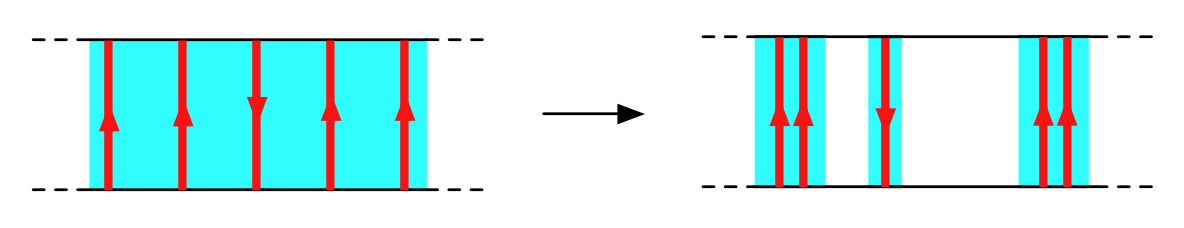}
\caption{\small In weakly coupled plasmas parallel currents tend to cluster together.}
\label{redistribution3}
\end{center}
\end{figure}

Somewhat pictorially, the main similarity is the tendency to `filamentation'. Very roughly  speaking, in weakly coupled plasmas one can think of this as the tendency of similarly oriented currents to cluster together  \cite{Weibel:1959zz,Mrowczynski:1988dz,Mrowczynski:1993qm,Mrowczynski:1994xv,Mrowczynski:1996vh} (see e.g.~\cite{Arnold:2003rq} for a more precise discussion), as sketched in Fig.~\ref{redistribution3}.
We emphasize however that there are also obvious differences. In a weakly coupled plasma the anisotropy is `dynamical' in the sense that it arises from the momentum distribution of the particles that compose the plasma, and it will disappear as the plasma eventually isotropizes and thermalizes. In contrast, in our case the plasma is static and intrinsically anisotropic because of the presence of dissolved extended objects. Yet the existence of instabilities in weakly coupled plasmas may be discovered by splitting the degrees of freedom into `hard' and `soft' modes \cite{Mrowczynski:1993qm,Mrowczynski:1994xv,Randrup:2003cw,Romatschke:2003ms,Arnold:2003rq,Romatschke:2004jh,Arnold:2004ti,Rebhan:2004ur,Arnold:2005vb}, where the former carry momenta of order $T$ and the latter of order  $\gym T$.  The hard modes  are then treated as particles whose distribution function $f(t,\mathbf{x},\mathbf{p})$ is often taken to be  static (over an appropriate time scale) and homogeneous in position space but anisotropic in momentum space,  
$f(t,\mathbf{x},\mathbf{p})= f_\mt{aniso}(\mathbf{p})$.  The soft degrees of freedom are then treated as fields in this `hard background', and their dispersion relations reveal some unstable modes. Although in strongly coupled plasmas with a gravity dual there is in general no quasi-particle description and no separation between hard and soft degrees of freedom (see e.g.~\cite{review} and references therein), our gravity solution may be viewed in a similar spirit to the extent that it provides a homogeneous, static, anisotropic and unstable background (for certain values of $(a,T)$). 

It would be interesting to explore whether this analogy can be made more precise. A first necessary step would consist of a  comparison of the degree of anisotropy in our solution and that of weakly coupled plasmas. In the description of the latter one often considers an anisotropic distribution that can be obtained by stretching or contracting an isotropic distribution \cite{Romatschke:2003ms,Romatschke:2004jh}, i.e.~one chooses
\be
f_\mt{aniso}(\mathbf{p}) 
\propto
 f_\mt{iso}\left( \sqrt{\mathbf{p}^2 + 
\xi (\mathbf{p}\cdot \mathbf{n})^2} \right) \,,
\ee
where $\mathbf{n}$ is a unit vector in the direction of the anisotropy (the $z$-direction in our case), 
\be
\xi = \frac{\langle p_T^2 \rangle}{2\langle p_L^2 \rangle} - 1 
\ee
is an adjustable parameter specifying the degree of anisotropy, and  
$\langle p_T^2 \rangle$ and $\langle p_L^2 \rangle$ are the average particle  momenta in the directions transverse and longitudinal to the anisotropy, respectively.  The isotropic case corresponds of course to $\xi=0$, and the initial stages of a heavy ion collision are expected to correspond to $\xi >0$ \cite{early}. In order to compare with the anisotropy of our solution, we note that $\xi$ is related to the anisotropy defined in terms of the pressure difference  
\be
\Delta = \frac{P_{xy}}{P_z} -1 
\ee
through \cite{Martinez:2009mf}
\bea
\lim_{\xi \ra 0} \Delta &=& \frac{4}{5} \xi + O(\xi^2) \,, \\
\lim_{\xi \ra \infty} \Delta &=& \frac{1}{2} \xi + O(\sqrt{\xi}) \,.
\eea
It is interesting to note that in our case, if one considers at least the metastable homogeneous phase, then  $\Delta$ actually spans the entire range $\Delta \in (-\infty,\infty)$, part of which ($\Delta < -1$) cannot be mapped to any value of $\xi \geq -1$. The simplest way to see this is to  consider approaching the blue line in the phase diagram of Fig.~\ref{phasediagram} along the $T=0$ axis. At the point $(T,a_\mt{$P_z$})$ we have $P_z=0$ and, from eqn.~\eqn{zero}, $P_{xy}>0$. Thus as we approach the curve from the left $\Delta \ra -\infty$, whereas as we approach it from the right 
$\Delta \ra \infty$. We also note from eqns.~\eqn{Pxyhigh} and \eqn{Pzhigh} that for a small anisotropy ($T \gg a$)
\be
\xi \simeq \frac{5}{4} \Delta \simeq \frac{5a^2}{16 \pi^2 T^2} > 0 \,,
\ee
whereas on the magenta curve we have $\Phi=0$ and therefore, from eqn.~\eqn{deltap}, $\Delta=\xi=0$. 

It is interesting to mention that a set of intrinsically anisotropic hydrodynamic equations, possibly including dissipation, can be written down for the hard modes  \cite{ani,ani2,ani3,ani4,ani5,ani6,ani7,ani8,ani9}. These equations can be derived, for example, by taking moments of the Boltzmann equation with an anisotropic distribution function $f(t,\mathbf{x},\mathbf{p})$ \cite{ani5,ani7}, and might provide an effective description of the system formed in the ultrarelativistic collision of two heavy ions in the early stage $\tau < \tau_\mt{iso}$ after the collision in which the anisotropy is large. In these equations the magnitude of the anisotropy $\xi$ is allowed to vary over space and time, but the orientation of the anisotropic direction is fixed. In the absence of dissipation, the generalized equations in which fluctuations of the anisotropy direction are also allowed have been written down in \cite{blackfold4}. In the latter case the anisotropy is induced by the presence of a $p$-dimensional charge density whose local orientation determines the anisotropic direction. 
Our solution may be viewed as the static (thermodynamic) limit of such a hydrodynamic theory with $p=2$. Perturbing our gravity solution and applying the fluid/gravity correspondence along the lines of \cite{fluidgravity6} it should be possible to derive the corresponding anisotropic hydrodynamic equations to (in principle) any desired order in the derivative expansion. Of course, the form of the dissipative terms at every order in the expansion can be written down on general grounds, but the derivation based on perturbations of our gravity solution would also predict specific values of the hydrodynamic coefficient of each term, as in \cite{Baier:2007ix,fluidgravity6,makoto}. Note that, unlike in Ref.~\cite{Baier:2007ix}, in our case the anomaly already plays a role in the thermodynamics of the system, and therefore it will presumably also play a role in the hydrodynamics even at the lowest order. 

The nature of some of the instabilities of our solution is also reminiscent of  the phenomenon of cavitation. In an ordinary flowing liquid, cavitation is the formation of bubbles of vapour in regions in which the pressure of the liquid drops below its vapour pressure \cite{cavitation}. Cavitation has been proposed \cite{torri,torri2,torri3} as a mechanism that would lead to clustering and fragmentation of the QGP into droplets that would subsequently evaporate, thus providing a new scenario for how hadronization is achieved. In this context the analogue of vapour pressure is played by the pressure of the vacuum, $P=0$, so cavitation would presumably occur if viscosity corrections to the pressure of the plasma become large enough to drive it negative. At that point bubbles of vacuum can form causing a fragmentation of the plasma ball. The possibility that this takes place in the QGP has been investigated in \cite{torri3,inv,inv2}.
The result depends of course on the values assumed for the transport coefficients of the QGP, in particular on the bulk and shear viscosities. A study using holographically determined coefficients has appeared recently \cite{holo}. Returning to our solution, we see that the role of the vapour pressure is played by the pressure of the isotropic phase: the homogeneous anisotropic phase becomes unstable against the formation of isotropic bubbles precisely when the anisotropic pressure drops below the pressure of the isotropic phase at the same temperature (eqn.~\eqn{pcondition}). We emphasize, however, that in the case of the mechanism proposed in \cite{torri,torri2,torri3} the pressure drop is due to a dynamical effect, namely to the viscosity corrections to the thermodynamic pressure that result from the expansion of the plasma. In contrast, in our case this is a static effect presumably resulting from the interaction of the D7-branes dissolved in the plasma. Yet, the similarity is intriguing and it would be interesting to investigate whether our solution can be viewed as a static toy-model for the dynamical situation. 

Before closing our discussion of instabilities, we explain why an instability studied in \cite{ALT} does not directly apply to our setup. The authors of \cite{ALT} analyzed the spectrum of supergravity fluctuations at zero temperature around the Lifshitz geometry \eqn{metric_ALT} (in the Einstein frame). By KK reducing on the $S^5$, they determined the mass spectrum of the different modes, and by analyzing the corresponding Laplace-like equations in the metric  \eqn{metric_ALT}, they conjectured a stability bound on the values of the masses, which can be thought of as a generalization of the usual AdS Breitenlohner-Freedman (BF) bound \cite{BF,BF2,BF3,BF4}. They concluded that a few modes violate the bound, indicating an instability. The key point is that this analysis applies only when the solution is exactly Lifshitz everywhere, but ceases to apply once this solution is UV completed with an AdS$_5$ geometry. In fact, the modes identified in \cite{ALT} as unstable with respect to the `Lifshitz BF bound' are stable with respect to the AdS BF bound.  Yet, it would be extremely interesting to investigate the remaining possibility that these modes `want to condense' in the IR, where they are effectively unstable, in a way similar to that observed in the condensation of neutral scalar fields \cite{super3} in holographic superconductors \cite{super,super2,super3}. Since the modes in question possess angular momentum on the $S^5$, this would lead to a spontaneous breaking of the $SO(6)$ symmetry of our solution at low temperatures.

\begin{figure}[tbp]
\begin{center}
\includegraphics[width=0.7\textwidth]{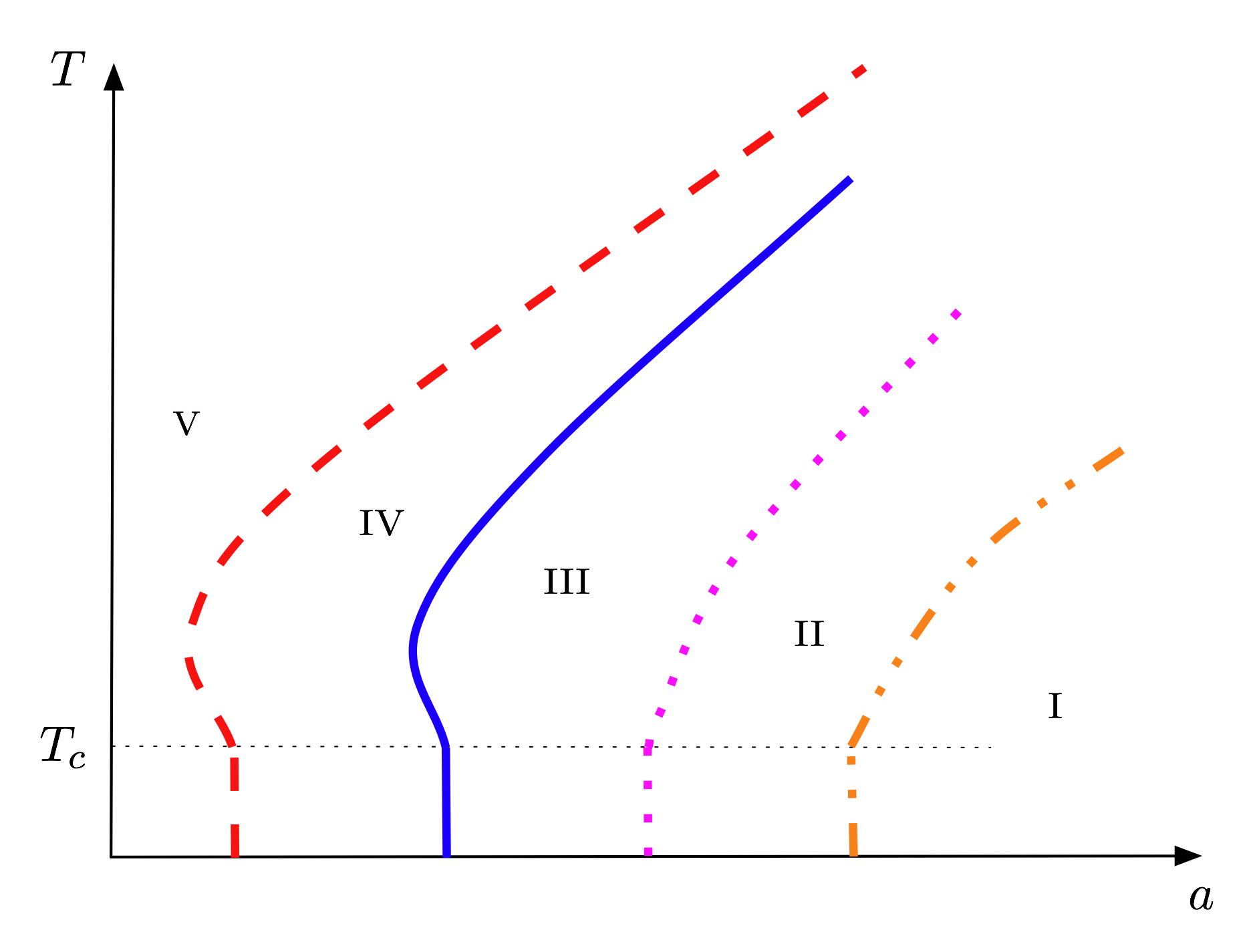}
\caption{\small Modification of the phase diagram at temperatures $T<1/L$ if the $x$- or $y$-directions are periodically identified with period $L$. }
\label{phasediagram2}
\end{center}
\end{figure}

Throughout this paper we have implicitly assumed that the $xy$-directions are not compact. Suppose now that one of them (say $x$) is periodically identified with period $L$. 
If periodic boundary conditions are imposed on the fermions around the circle then the physics remains as we have described above. However, if antiperiodic boundary conditions are imposed, then we know \cite{witten} that the undeformed ${\cal N}=4$ SYM theory undergoes a thermal phase transition at a critical temperature $T_c=1/L$  between a confined and a deconfined phase.   In the gravity description this corresponds to a Hawking-Page-like phase transition \cite{HP} between geometries without and with a black hole, respectively. This transition extends to all values of $a$ at exactly the same critical temperature because the Euclidean metrics in the two phases are related by a simple relabeling of the $t, x$ coordinates. The high-temperature geometry is dual to the deconfined phase and is given by \eqn{sol1}, whereas the low-temperature metric describes the confining phase and is given by the double analytic continuation of \eqn{sol1}:
\be
ds^2 =  \frac{e^{-\frac{1}{2}\phi}}{u^2}
\left( -dt^2+ \cf \cb\, dx^2+dy^2+ \ch dz^2 +\frac{ du^2}{\cf}\right) \,.
\label{sol3}
\ee
Clearly, the Euclidean metrics corresponding to the Lorentzian metrics \eqn{sol1} and \eqn{sol3} are related by the simple exchange 
$t \leftrightarrow x$ (cf.~\cite{QHE}). Since the free energy density is simply obtained by evaluating the Euclidean action on the corresponding Euclidean solution, it follows that the free energy density $\tilde F$ in the confining phase is obtained from the free energy density in the deconfined phase $F$ by exchanging $\beta \leftrightarrow L$. By symmetry, this implies that the phase transition must happen exactly at the point  
$L=\beta$ regardless of the value of $a$. At leading order in the large-$\nc$ expansion $F$ is independent of $L$,\footnote{The masses of KK states around the $x$-circle depend on $L$, and so these states will give an $L$-dependent but $\nc$-subleading contribution when loop effects are considered.} and therefore $\tilde F$ and its derivatives $\tilde \Phi = \partial \tilde F / \partial a$ and $\partial^2 \tilde F / \partial a^2$ are temperature-independent.\footnote{Incidentally, note that this immediately implies that the entropy density vanishes at leading order in the confined phase, $\tilde s=\partial \tilde F / \partial T=0$, which is consistent with the absence of a horizon in the metric \eqn{sol3}.} The same is true for the pressure in the $z$-direction $\tilde P_z$, which is obtained from $P_z$ 
by the exchange $\beta \leftrightarrow L$. Since the free energy, its $a$-derivatives and the $z$-pressure are continuous across the phase transition (though their $T$-derivatives are not), we conclude that the phase diagram in the $(a,T)$-plane  `freezes out' below $T_c$, meaning that the four curves in the diagram extend from $T=T_c$ to $T=0$ in a temperature-independent way, as shown in Fig.~\ref{phasediagram2}. 

We plan to report elsewhere on the effects of the anisotropy on several observables. For example, anisotropy effects on heavy-quark energy loss in a weakly coupled plasma have been studied in \cite{loss,loss2,loss3}, and effects on momentum broadening, in particular on the jet quenching coefficient $\hat q$, in \cite{quench}. In the isotropic case, the drag force on a heavy quark in a strongly coupled plasma has been considered in \cite{drag,drag2,drag3,drag4,drag5,drag6,drag7,drag8,drag9}, its energy loss through Cherenkov-emission of heavy quarkonium mesons in \cite{cherenkov,cherenkov2}, and jet quenching in \cite{jet,jet2,jet3,jet4,jet5,jet6,jet7,jet8}. Recently the drag force was also studied in an anisotropic background in \cite{anidrag} and in a Lifshitz background in \cite{Fadafan:2009an}. Extending the calculations of these references to our anisotropic solution will allow a comparison with the results for weakly coupled anisotropic plasmas. 

Work is also in progress on the anisotropy effects on heavy quarkonium physics, which at weak coupling have been studied in \cite{meson,meson2,meson3,meson4}. In our context, consideration of dynamical quarks and mesons requires the introduction of flavour branes \cite{flavour} wrapping an $S^3$ inside the $S^5$. The meson spectrum and the meson dispersion relations can be calculated by extending the analysis of \cite{flavour2,back2,dispersion}. It is also  interesting to understand the effects of the anisotropy on the meson dissociation phase transition uncovered in \cite{melting,back2,melting2,melting3,melting4}. One conclusion that can be anticipated is the fact that the in-medium meson limiting velocities \cite{back2} in the $xy$- and in the $z$-directions will differ, since these are  respectively given by   
\be
v_{xy}=\sqrt{-\frac{g_{tt}}{g_{xx}}} \qquad \mbox{and} \qquad  
v_{z}=\sqrt{-\frac{g_{tt}}{g_{zz}}} 
\ee
evaluated at the tip of the flavour D7-branes at the point at which the dissociation transition takes place.  

A third set of observables which can be extracted from our solution is that related to photon and dilepton thermal production by the plasma. At weak coupling these have been calculated both in the absence \cite{photonprod1,photonprod2,photonprod3} (see \cite{photonprod0} for a review) and in the presence of anisotropy \cite{photon,dilep,dilep2,dilep3}. Corresponding studies at strong coupling in the isotropic case include \cite{photonprod4,photonprod5,photonprod6,photonprod7}, and we are currently  extending these calculations to the anisotropic case.

\section*{Acknowledgements}
We thank A.~Buchel, J.~Casalderrey-Solana, B.~Fiol, J.~Garriga, A.~Hashimoto, E.~Kiritsis, F.~Marchesano, D.~Marolf, M.~Martinez, G.~Moore, R.~Myers, I.~Papadimitriou, M.~Roberts, K.~Skenderis, and very specially R.~Emparan for useful discussions. We are particularly  grateful to I.~Papadimitriou for sharing Ref.~\cite{Yiannis} with us prior to publication. DT thanks the organizers and participants of the ``Great Lakes Strings 2011'' conference at the University of Chicago for creating a stimulating meeting, where this work was presented. We are supported by 2009-SGR-168, MEC FPA2010-20807-C02-01, MEC FPA2010-20807-C02-02 and CPAN CSD2007-00042 Consolider-Ingenio 2010 (DM), and by PHY04-56556, DE-FG02-91ER40618 and DE-FG02-95ER40896 (DT).

\appendix


\section{Derivation of the solution}
\label{derivation}

This appendix contains a detailed description of the procedure we have followed to derive our solution. 

The starting point is type IIB supergravity in the string frame (see for example \cite{Joesbigbook}). Following Ref.~\cite{ALT} (whose relevant details we review in Appendix~\ref{comparison}), we seek a solution in which only the metric, the axion, the dilaton and the RR five-form are non-trivial. The solution can be consistently derived from the equations of motion associated to the truncated action
\bea
S&=&\frac{1}{2 \kappa_\mt{10}^2}\int d^{10}x \, \sqrt{-g} \left[
e^{-2\phi}(R+4\partial_M\phi \partial^M\phi)
-\frac{1}{2}F_1^2-\frac{1}{4 \cdot 5!}F_5^2 
\right],
\label{10daction}
\eea
where $M=0,\ldots,9$ and $F_1=d\chi$ is the axion field-strength.

The dilaton equation of motion resulting from this action is
\be
R+4g^{MN}\left(\nabla_M \nabla_N \phi - \partial_M \phi \partial_N \phi\right)=0\,,
\label{dilatonEOM}
\ee
and the Einstein equations are
\begin{eqnarray}
  R_{MN}+2 \nabla_M\nabla_N \phi +\frac{1}{4} g_{MN}e^{2\phi}\partial_P \chi \partial^P \chi  -\frac{1}{2}e^{2\phi}\left(F_M F_N +\frac{1}{48} F_{MABCD}F_N^{\ ABCD}\right)=0.
 \label{EinsteinEOM}
\end{eqnarray}
The forms obey the following equations of motion, Bianchi identities, and self-duality constraint
\be
d\star F_1=0=d\star F_5\sac dF_1=0=dF_5\sac \star F_5=F_5\,,
\label{forms}
\ee
where $\star$ is the ten-dimensional Hodge dual.
To model an anisotropic plasma with an asymmetry between, say, the $xy$- and the $z$-directions, we assume the following Ansatz for the string-frame metric
\bea
ds^2 =
 \frac{1}{u^2}\left( -\cf \cb\, dt^2+dx^2+dy^2+\ch dz^2 +\frac{ du^2}{\cf}\right)+ {\cal Z} \, d\Omega^2_{S^5}\,.
 \label{10dmetric}
\eea
Without loss of generality (see Section \ref{action-section}), in this equation we have set to unity an overall constant $L$ with dimensions of length.  $\cf$, $\cb$, $\ch$, and ${\cal Z}$, as well as the dilaton $\phi$, are taken to be functions of the radial coordinate $u$ only. $\Omega_{S^5}$ is the metric of a unit five-sphere (or of any other Einstein five-manifold). In this parametrization the boundary is located at $u=0$. As explained in the main text, we have used reparametrization invariance to fix $g_{xx}$ and $g_{yy}$ as in (\ref{10dmetric}). Having done this, we cannot eliminate $\cb$ in general, although we still have a scaling symmetry in the time coordinate 
that can be used to set $\cb_\mt{bdry}=1$. (Here and in the main text we use the subscript `bdry' to denote the value of a field at $u=0$.) Similarly, there is a scaling symmetry in $z$ that we can use to set $\ch_\mt{bdry}=1$. $\cf$ is a `blackening factor' that allows for the introduction of a black hole in the geometry. It vanishes at the position of the horizon, $u=\uh$.

We take the (magnetic part of the) RR five-form to be proportional to the volume form of the five-sphere 
\bea 
F_5=\alpha\, (\Omega_{S^5}+\star \Omega_{S^5})\,.
\label{F5}
\eea
Equations (\ref{dilatonEOM})-(\ref{EinsteinEOM}) imply that at the boundary we must have $\alpha = 4 \, e^{-\phi_\mt{bdry}}$. As explained in Section \ref{number}, we can choose $\phi_\mt{bdry}=0$ without loss of generality, which corresponds to  $\alpha=4$. Following \cite{ALT}, we choose the axion to be linear in the boundary coordinate $z$ 
\bea
\chi= a \, z\,,
\label{chi}
\eea
and we simplify our Ansatz further by setting
\be
\ch=e^{-\phi}\,, \qquad \qquad {\cal Z}=e^{\frac{1}{2}\phi}\,.
\label{simplifying_ansatz}
\ee
Clearly, (\ref{F5})-(\ref{chi}) satisfy (\ref{forms}).

With the choices above the sphere part of the space factorizes in Einstein frame, and the geometry becomes a direct product of the sphere and an asymptotically locally AdS space. We can therefore reduce the ten-dimensional action (\ref{10daction}) to the five-dimensional axion-dilaton AdS gravity in (\ref{action}), with the five-form flux giving rise to the cosmological constant. The metric (\ref{10dmetric}) also reduces (in Einstein frame) to (\ref{sol1}). 

Notice that the asymptotic equations of motion also imply that 
\bea
\cf_\mt{bdry}=1/{\cal Z}_\mt{bdry}\,.
\eea
With our choices $\phi_\mt{bdry}=0$ (which imposes $\cf_\mt{bdry}=\ch_\mt{bdry}={\cal Z}\mt{bdry}=1$) and $\cb_\mt{bdry}=1$, the metric (\ref{10dmetric}) induces on the boundary a flat metric in which the $t,x,y,z$ coordinates are canonically normalized. 


\subsection{Equations of motion}

We obtain now the explicit expressions for the equations of motion that we are ultimately going to solve. 

We start with some counting. The Einstein equations (\ref{EinsteinEOM}) are diagonal, as a consequence of the fact that the sphere is Einstein and that the warp factors in the ansatz depend only on one coordinate, $u$. We have  one equation from (\ref{dilatonEOM}) and five equations from (\ref{EinsteinEOM}), which are linearly independent (the $x$- and $y$-directions give one independent equation, and so do all the sphere-directions). Out of these equations, five turn out to be second-order and one is first-order, namely a constraint. We have only three unknown functions to solve for: $\phi$, $\cb$, and $\cf$.

What we did in practice is the following. The six independent Einstein and dilaton equations we started with read
\bea
&& 0=-\frac{e^{3\phi}}{2}a^2-\frac{8 e^{-\phi/2}}{u^2}+\frac{8 \cf}{u^2}-\frac{4 \cf \cb'}{u \cb}-\frac{\cf \cb'^2}{2 \cb^2}-\frac{5\cf'}{u}
\cr && \hskip 2cm +\frac{3 \cb'\cf'}{2\cb}+\frac{5\cf\phi'}{2u}-\frac{5\cf\cb'\phi'}{4\cb}-\frac{5\cf'\phi'}{4}+\frac{\cf\cb''}{\cb}+\cf''\,,
\label{eq0}
\\
&&\cr 
&&0=
\frac{e^{3\phi}u}{4\cf}a^2-\frac{4}{u}+\frac{4 e^{-\phi/2}}{u\cf}+\frac{\cb'}{2\cb}+\frac{\cf'}{\cf}-\frac{5\phi'}{4}\,,
\label{eq1}
\\
&&\cr 
&&0=-\frac{e^{3\phi}}{2\cf}a^2-\frac{8}{u^2}+\frac{8 e^{-\phi/2}}{u^2 \cf}+\frac{\cb'}{u\cb}+\frac{2\cf'}{u\cf}
-\frac{11 \phi'}{2 u}+\frac{\cb'\phi'}{2\cb}+\frac{\cf' \phi'}{\cf}-\frac{5 \phi'^2}{4}+\phi''\,,
\label{eq2}
\\
&&\cr 
&&0=\frac{e^{3\phi}}{\cf}a^2-\frac{16}{u^2}+\frac{16 e^{-\phi/2}}{u^2 \cf}+\frac{2\cb'}{u\cb}+\frac{\cb'^2}{\cb^2}+\frac{10 \cf'}{u\cf}
\cr && \hskip 2cm -\frac{3\cb' \cf'}{\cb\cf}+\frac{\phi'}{u}+\frac{5 \cf'\phi'}{2\cf}-\frac{9 \phi'^2}{4}-\frac{2 \cb''}{\cb}-\frac{2 \cf''}{\cf}+5\phi''\,,
\label{eq3}
\\
&&\cr
&&0=\frac{e^{3\phi}}{\cf}a^2+\frac{3\phi'}{u}-\frac{\cb'\phi'}{2\cb}-\frac{\cf'\phi'}{\cf}+\frac{5\phi'^2}{4}-\phi''\,,
\label{eq4}
\\
&&\cr
&& 0=\frac{20 e^{-\phi/2}}{u^2}-\frac{20 \cf}{u^2}+\frac{4 \cf \cb'}{u\cb}+\frac{\cf \cb'^2}{2 \cb^2}+\frac{8\cf'}{u}-\frac{3\cb' \cf'}{2\cb}
\cr && \hskip 2cm -\frac{17 \cf \phi'}{2u}+\frac{5 \cf \cb' \phi'}{4 \cb}+\frac{5 \cf'\phi'}{2}-\frac{17\cf\phi'^2}{8}-\frac{\cf\cb''}{\cb}+\frac{5\cf\phi''}{2}-\cf''
\label{eq5}
\,,
\eea
where primes denote derivatives with respect to $u$. We used the last two equations, (\ref{eq4})-(\ref{eq5}), to solve for $\phi''$ and $\cb''$ and plugged the results into the three other dynamical equations, (\ref{eq0}) and (\ref{eq2})-(\ref{eq3}), to obtain three constraints:
\bea
&& 0=2 e^{3\phi}a^2+\frac{12 e^{-\phi/2}}{u^2}-\frac{12 \cf}{u^2}+\frac{3 \cf'}{u}+\frac{3 \cf \phi'}{2 u}-\frac{5\cf \cb'\phi'}{4\cb}
-\frac{5\cf'\phi'}{4}+\cf\phi'^2\,,
\label{con1}
\\
&&\cr 
&&0=
\frac{e^{3\phi}u}{2\cf}a^2-\frac{8}{u}+\frac{8 e^{-\phi/2}}{u\cf}+\frac{\cb'}{\cb}+\frac{2 \cf'}{\cf}-\frac{5\phi'}{2}\,,
\label{con2}
\\
&&\cr 
&&0=\frac{e^{3\phi}}{\cf}a^2+\frac{24}{u^2}-\frac{24 e^{-\phi/2}}{u^2 \cf}-\frac{6 \cb'}{u\cb}-\frac{6\cf'}{u\cf}
+\frac{18 \phi'}{u}-\frac{5 \cb'\phi'}{2\cb}-\frac{5 \cf' \phi'}{\cf}+2\phi'^2\,.
\label{con3}
\eea
Only two of the four constraints (\ref{eq1}) and (\ref{con1})-(\ref{con3}) are independent. We can use for example (\ref{con1}) and (\ref{con2}) to solve for $\cf'$ and $\cb'$. Plugging the solution for $\cf'$ into (\ref{eq4}) one obtains an {\it algebraic} equation for $\cf$:
\bea
\cf=\frac{e^{-\frac{1}{2}\phi}}{4(\phi'+u\phi'')}\left(a^2 e^{\frac{7}{2}\phi}(4u+u^2\phi')+16\phi'\right)\,.
\label{eq_F_app}
\eea
Using this expression in the solution for $\cb'$ obtained from (\ref{con1}) and (\ref{con2}), one arrives at the first-order equation for $\cb$
\bea
\frac{\cb'}{\cb}= \frac{1}{24+10 u\phi'}\left(24\phi'-9u\phi'^2+20u\phi''\right)\,.
\label{eq_B_app}
\eea
This equation determines $\cb$ up to an overall multiplicative constant. This is of course a result of the aforementioned scaling symmetry in $t$, which allows us to set $\cb_\mt{bdry}=1$.

Combining these results with the solution for $\phi''$ obtained from (\ref{eq4})-(\ref{eq5}) we can isolate a third-order equation for the dilaton alone:
\bea
0&=&\frac{256 \phi ' \phi ''-16 \phi '^3
   \left(7 u \phi '+32\right)}{u\,  a^2 e^{\frac{7 \phi }{2}}
   \left(u \phi '+4\right)+16 \phi '} +\frac{\phi ' }{u \left(5 u \phi '+12\right) \left(u \phi
   ''+\phi '\right)}\times \cr &&  \times \Big[13 u^3 \phi '^4+8 u \left(11
   u^2 \phi ''^2-60\phi''-12 u \phi '''\right)+u^2 \phi '^3
   \left(13 u^2 \phi ''+96\right) \cr &&\hskip .3cm  +2 u \phi '^2 \left(-5 u^3 \phi
   '''+53 u^2 \phi ''+36\right)+\phi ' \left(30 u^4 \phi
   ''^2-64 u^3 \phi '''-288+32 u^2 \phi
   ''\right)\Big]\,.
   \label{3order_dil}
   \cr && 
\eea
At this point, we observe that the anisotropy parameter $a$ and the undifferentiated dilaton only enter in eqn.~(\ref{3order_dil}) through the combination $ a^2 e^{\frac{7}{2}\phi}$, so that we can shift the dilaton
\be
\phi \to \tilde\phi\equiv \phi+\frac{4}{7}\log a\,,
\label{dilaton_shifted}
\ee
and eliminate $a$ from (\ref{3order_dil}) altogether. The only place where $a$ will appear explicitly is as an overall factor of $a^{2/7}$ in the eqn.~(\ref{eq_F_app}) for $\cf$. We also note that, since we fixed $\phi_\mt{bdry}=0$, we will have
\be
a = e^{\frac{7}{4}\tilde\phi_\mt{bdry}}\,.
\label{beta_app}
\ee
Solutions with different $\tilde\phi_\mt{bdry}$ correspond then to systems with different anisotropies.

We can solve the equations of motion above analytically in the limiting cases of high and low temperature (see Appendices \ref{highApp} and \ref{lowApp}), but for intermediate regimes we have to resort to numerical integration. In the next section we give some details about the procedure we have followed.


\subsection{Numerics}

We have solved eqn.~(\ref{3order_dil}) numerically (after having shifted the dilaton as in  (\ref{dilaton_shifted})), integrating from the horizon $u=\uh$ to the boundary $u=0$.\footnote{In practice, one actually integrates from $\uh-\epsilon$ to $\epsilon$, where $\epsilon\ll 1$. } We expand the dilaton as 
\be
\tilde\phi =\tilde\phi_\mt{H}+ \sum_{n\ge 1} \tilde\phi_n \, (u-u_H)^n \,,
\label{expansion_phi}
\ee
and similarly for the other fields, with the only restriction that  
$\cf_\mt{H}=0$. Inserting the expansion \eqn{expansion_phi} in (\ref{3order_dil}) determines all the dilaton coefficients in terms of $\tilde\phi_\mt{H}$ and $u_H$. These in turn determine the coefficients of $\cf$ and $\cb$ through eqns.~(\ref{eq_F_app})-(\ref{eq_B_app}). 
Alternatively, one can substitute all the expansions into eqs.~(\ref{eq0})-(\ref{eq5}), and then solve at the same time for all the coefficients order by order. Of course, the two procedures yield the same result.  Here we just report the expressions for a few of the coefficients:
\bea
\tilde\phi_1 &=& -\frac{4 e^{\frac{7}{2}\tilde\phi_\mt{H}}\uh}{16+e^{\frac{7}{2}\tilde\phi_\mt{H}}\uh^2}\,,\qquad \qquad \tilde\phi_2 = \frac{2 e^{7\tilde\phi_\mt{H}}\uh^2\left(128+e^{\frac{7}{2}\tilde\phi_\mt{H}}\uh^2\right)}{\left(16+e^{\frac{7}{2}\tilde\phi_\mt{H}}\uh^2\right)^3}\,,
\cr
\tilde\phi_3 &=& -\frac{4e^{\frac{7}{2}\tilde\phi_\mt{H}}\left(-98304+34816 e^{\frac{7}{2}\tilde\phi_\mt{H}} \uh^2
+52864 e^{7\tilde\phi_\mt{H}}\uh^4-336 e^{\frac{21}{2}\tilde\phi_\mt{H}}\uh^6+3e^{14\tilde\phi_\mt{H}}\uh^8\right)}{9\uh \left(16+e^{\frac{7}{2}\tilde\phi_\mt{H}}\uh^2\right)^5}\,.\cr &&
\eea
In the following we will also need the explicit expression for $\cf_1$:
\be
\cf_1 =  \frac{a^{\frac{2}{7}} e^{-\frac{1}{2}\tilde\phi_\mt{H}}}{4 \uh}
\left( 16+\uh^2 \, e^{\frac{7}{2}\tilde\phi_\mt{H}} \right) \,.
\label{f1}
\ee
After integrating (\ref{3order_dil}) numerically we use $\tilde\phi$ to  obtain $\cf$ and $\cb$ through (\ref{eq_F_app})-(\ref{eq_B_app}).


\subsection{Parameters}

We have seen that the solution is determined by two parameters, $\tilde\phi_\mt{H}$ and $\uh$. This was to be expected, since these determine the two physical parameters that the solution must depend on: the temperature and the anisotropy. As far as we have been able to verify with our numerical results, the map 
\be
T=T(\tilde\phi_\mt{H},\, \uh)\,,\qquad\qquad  a=a(\tilde\phi_\mt{H},\, \uh)
\label{map}
\ee
is one-to-one. 

The parameter $a$ can be easily extracted from the boundary value of the rescaled dilaton $\tilde\phi_\mt{bdry}$ through (\ref{beta_app}). To compute the temperature, we have already noted that the Euclidean continuation of the metric \eqn{sol1} in the $(t_\mt{E},u)$-plane near $\uh$ takes the form (\ref{euclideanized}).
In terms of new coordinates
\be
\vartheta = \frac{\cf_1 \sqrt{\cb_\mt{H}}}{2}\, t_\mt{E} \sac 
\rho= \frac{2 e^{-\frac{1}{4}\phi_0}}{\uh \sqrt{\cf_1}} \sqrt{u-\uh}\,,
\ee
this piece of the metric becomes just $ds^2_\mt{E} \approx d\rho^2 + \rho^2 d\vartheta^2$. Regularity at $u=\uh$ then requires that $\vartheta$ have period $\delta \vartheta = 2\pi$, and therefore that 
$\delta t_\mt{E} = 4 \pi / \cf_1 \sqrt{\Bh}$. We identify this with the inverse temperature:
\bea
T = \frac{1}{\delta t_\mt{E}}  =  
\sqrt{\Bh}\, \frac{e^{\frac{1}{2} (\phit_\mt{bdry} - \phith)}}{16 \pi  \uh} 
\left( 16+\uh^2 \, e^{\frac{7}{2} \phith} \right) \,,
\eea
where in the last step we used \eqn{beta_app} and \eqn{f1}.

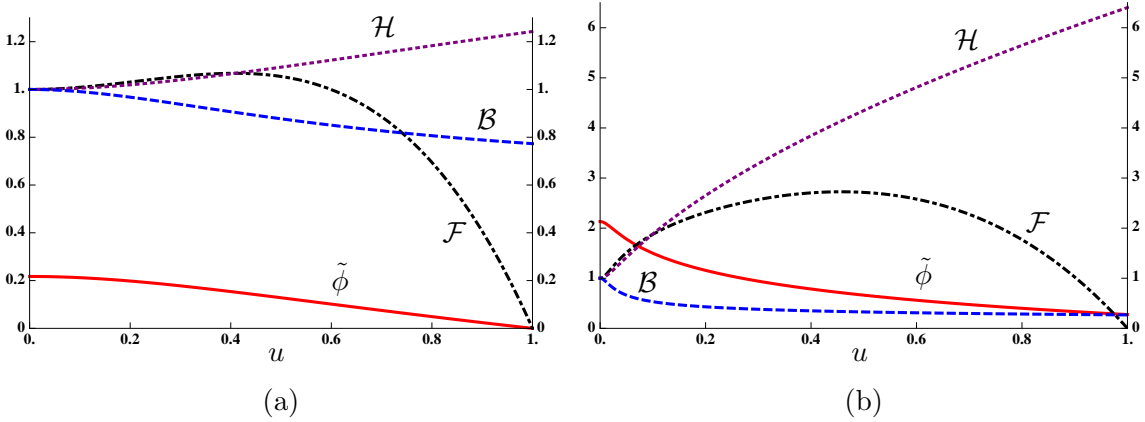
\begin{figure}[tb]
\begin{center}
\begin{tabular}{cc}
\includegraphics[scale=0.8]{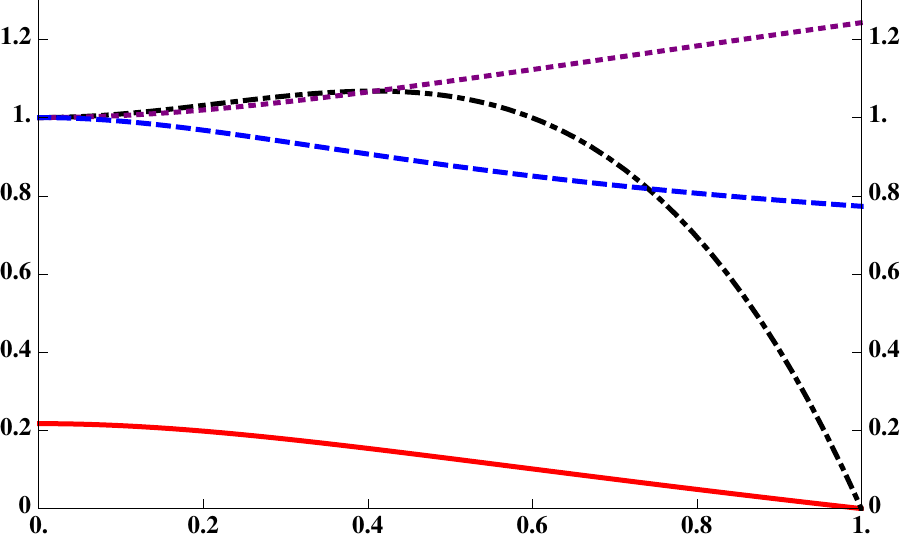}
\put(-109,-5){$u$}
\put(-109,-14){$$}
\put(-85,22){$\tilde\phi$}
\put(-70,118){$\ch$}
\put(-30,83){$\cb$}
\put(-43,40){$\cf$}
&
\includegraphics[scale=0.8]{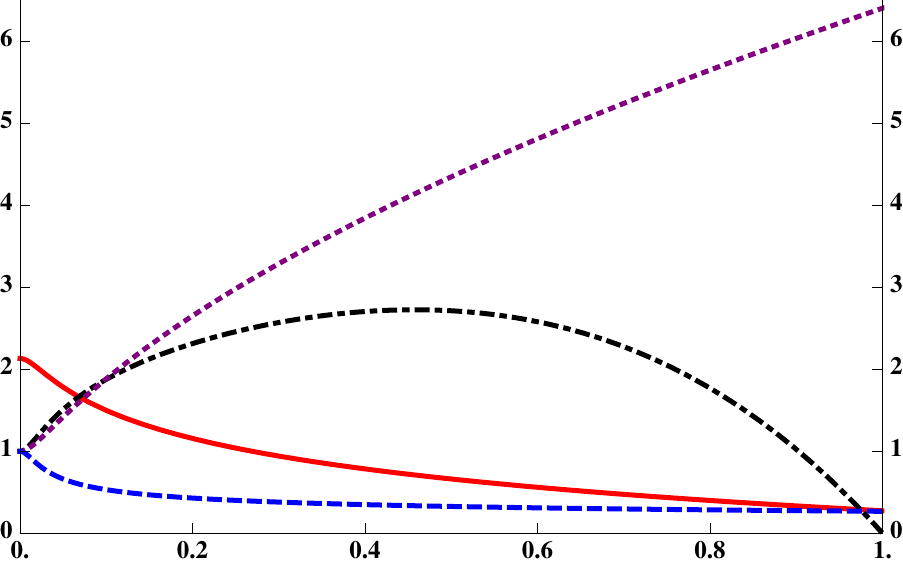}
\put(-109,-5){$u$}
\put(-109,-14){$$}
\put(-85,24){$\tilde\phi$}
\put(-70,113){$\ch$}
\put(-190,21){$\cb$}
\put(-43,44){$\cf$}
\\
(a) & (b)
\end{tabular}
\caption{\small (a) The metric functions for $a/T\simeq 4.4$, with $\uh=1$ and $\tilde\phi_\mt{H}=0$. (b) The metric functions for $a/T\simeq 86$, with $\uh=1$ and $\tilde\phi_\mt{H}=0.275$.
       \label{plots}
   }
 \end{center}
 \end{figure}

By repeating the integration procedure for different initial values of $\tilde\phi_\mt{H}$ and $\uh$, we can scan the $(T,\, a)$ plane and construct the map (\ref{map}). Roughly, we find that $a$ depends  strongly on  $\tilde\phi_\mt{H}$, whereas $T$ depends  strongly on  $\uh$. To illustrate what the solutions look like,  we present in 
Fig.~\ref{plots}  the plots corresponding to two different initial conditions: $(\tilde\phi_\mt{H}=0,\, \uh=1)$, which gives a moderate anisotropy-to-temperature ratio $a/T\simeq 4.4$, and  $(\tilde\phi_\mt{H}=0.275,\, \uh=1)$, which gives instead $a/T\simeq 86$. In both cases we have checked that the metric component $g_{tt}$ in the Einstein frame is monotonically increasing towards the boundary.


\section{The IR Lifshitz solution}
\label{comparison}

In this paper we have studied the finite-temperature generalization \cite{prl} of the RG flow found in \cite{ALT} between an AdS$_5$ geometry in the UV and a Lifshitz-like anisotropic geometry in the IR. Neither the authors of Ref.~\cite{ALT} nor we have been able to write down the solution describing the entire RG flow analytically. However, the geometry at the IR endpoint of the flow, including its finite-temperature version, was found analytically in \cite{ALT}.  In this appendix we review this solution and compare the  parametrization of \cite{ALT} with our Ansatz (\ref{10dmetric}).

From Sections 2.3 and 2.5 of \cite{ALT} we see that the metric in string frame for the finite-temperature Lifshitz-like fixed point reads (with some obvious relabelings)
\bea
ds^2_s=\tilde R^2_s \left[r^{\frac{7}{3}}\left( -f(r)dt^2+dx^2+dy^2 \right)+r^\frac{5}{3}dz^2+\frac{dr^2}{r^\frac{5}{3}f(r)}\right]+R^2_s r^\frac{1}{3}d\Omega_{S^5}^2\,,
\label{metric_ALT}
\eea
where the subscript `s' stands for string frame. The radii, the blackening factor and the dilaton are given by 
\be
\tilde R^2_s = \frac{11}{12} R^2_s \sac 
f(r)= 1- \left(\frac{r_\mt{H}}{r}\right)^{11/3} \sac
e^{\phi(r)}=r^{2/3}e^{\phi_0} \,,
\label{formal}
\ee
where $r_\mt{H}$ and $\phi_0$ are arbitrary constants. The forms are as in eqs.~(\ref{F5})-(\ref{chi}). As anticipated above, if $r_\mt{H}=0$ then the metric \eqn{metric_ALT} in the Einstein frame is invariant under the anisotropic scaling 
\be
\{ t,x,y\} \to k \{ t,x,y\} \sac z \ra k^{2/3} z \sac r \ra k^{-1} r \,.
\ee
Note however that the dilaton and the forms are not invariant.

The coordinates inside the square brackets in the metric \eqn{metric_ALT} are dimensionless. Comparison with our parameterization is facilitated by making them dimensionful through the rescalings 
\be
\{ t,x,y \} \to \tilde R_s^{-1}R_s^{-7/6} \{ t,x,y \} \sac
z \to \tilde R_s^{-1}R_s^{-5/6}z \sac 
r \to \tilde R_s^{-6}R_s^{5} \,r \,.
\ee
These bring (\ref{metric_ALT}) to the form
\be
ds^2_{s} = \left( \frac{r}{R_s} \right)^{7/3} \Big( -f(r) dt^2 + dx^2 +dy^2 \Big) +
\left( \frac{r}{R_s} \right)^{5/3} dz^2 +  \left( \frac{R_s}{r} \right)^{5/3} \frac{dr^2}{f(r)} + 
L_s^{\frac{5}{3}} r^{\frac{1}{3}} d\Omega_{S^5}^2 \,,
\label{metric_ALT_2}
\ee
where $L_s=  \left(12/11\right)^{3/5} R_s$ and we have also rescaled $r_\mt{H} \to \left(11/12\right)^{3} r_\mt{H} R_s$, so that the blackening factor still takes the form given in \eqn{formal}. 

The relation between $r$ and the radial coordinate $u$ that we have used throughout this paper is given by
\be
\left( \frac{r}{R_s} \right)^{7/3} = \left( \frac{L_s}{u}\right)^2\,.
\ee
It is immediate to see that in terms of $u$ the metric (\ref{metric_ALT_2}) takes precisely the form (\ref{10dmetric}):\footnote{In (\ref{10dmetric}) we have set $L_s=1$.}
\be
ds^2_{s} = \frac{L_s^2}{u^2} \, \Big( -\cf \cb\, dt^2 + dx^2 +dy^2 
+ \ch \, dz^2 + \frac{du^2}{\cf} \Big) + 
L_s^2 \, {\cal Z} \, d\Omega_{S^5}^2 \,,
\ee
with
\bea
\cf &=& \frac{49}{36}\left(\frac{12}{11}\right)^{6/5}\left(\frac{u}{L_s}\right)^{2/7} \left(1- \frac{u^{22/7}}{u^{22/7}_\mt{H}} \right) 
 \,, \qquad \qquad 
 \cb = \frac{36}{49} \left(\frac{11}{12}\right)^{6/5} \left(\frac{L_s}{u}\right)^{2/7}\,, \cr
\ch& = & \left(\frac{L_s}{u}\right)^{10/7}\,, \qquad \qquad 
 {\cal Z} = \left(\frac{11}{12}\right)^{1/5}\left(\frac{L_s}{u}\right)^{2/7}\,,
\eea
and a dilaton given by
\bea
e^\phi &=& e^{\phi_0}\left(\frac{11}{12}\right)^{2/5} \left(\frac{L_s}{u}\right)^{4/7}\,.
\eea

By computing the area of the horizon, it was shown in \cite{ALT} that the entropy density associated with the metric (\ref{metric_ALT_2}) is proportional to $\nc^2 T^{8/3}$. By dimensional analysis, this implies that 
\be
s = \tilde c_\mt{ent}\nc^2 a^{1/3} T^{8/3}\,,
\label{sss}
\ee
where $\tilde c_\mt{ent}$ is a constant. Note however that the constant $\tilde c_\mt{ent}$ appearing above and the constant $c_\mt{ent}$ appearing in (\ref{slif}) need not be the same, as emphasized in \cite{interpolation}. The reason is that this constants can be rescaled by a rescaling of the $x,y,z$ coordinates. In the case that the solution approaches AdS$_5$ asymptotically, the normalization of these coordinates is fixed by requiring that the gauge theory metric at the boundary be canonically normalized. This normalization need not coincide with the normalization chosen for these coordinates in the case in which the solution is asymptotically Lifshitz, as is the IR geometry described by the metric \eqn{metric_ALT}.

If $r_\mt{H}=0$ then the scalar curvature of the string-frame metric \eqn{metric_ALT} diverges as $r^{-1/3}$ as \mbox{$r\to 0$} (and the same is true for the five-dimensional metric inside the square brackets). In contrast, all curvature invariants of the corresponding Einstein-frame metric are finite everywhere. Nevertheless, as pointed out in \cite{tidal} and elaborated upon in \cite{tidal2}, the hypersurface \mbox{$r=0$} is still problematic because tidal forces become infinite as this hypersurface is approached. Fortunately, if $r_\mt{H}>0$ then this hypersurface is hidden behind the horizon and the geometry is regular everywhere on and outside the horizon.


\section{Relation between free energies and pressures}
\label{FPrelations}

In this section it will be useful to work with total energies and charges, as opposed to the densities used in the main text. For this purpose we consider a box of sides $L_x, L_y$ and $L_z$ and denote the total quantities with a bar. Setting $V=L_x L_y L_z$ we have: 
\be
E = \frac{\bar E}{V} \sac s = \frac{\bar s}{V}  \sac 
F = \frac{\bar F}{V} \sac G = \frac{\bar G}{V} \sac 
\Phi = \frac{\bar \Phi}{L_x L_y} \sac a = \frac{\bar a}{L_z} \,.
\label{global}
\ee
Note that, as explained in Section \ref{sec-thermodynamics}, $a$ has dimensions of 1/length because it measures the number of D7-branes per unit length in the $z$-direction, whereas $\Phi$ has dimensions of energy/length$^2$ because it measures the energy cost per unit $xy$-area of introducing an additional D7-brane extending in the $xy$-directions. 

The first law in terms of the total charges and allowing for independent changes in the lengths of the box takes the form
\be
d\bar E= T d\bar s - L_x L_y P_z \, dL_z - L_y L_z P_x \, dL_x 
- L_z L_x P_y \, dL_y + \bar \Phi \, d\bar a \,,
\ee
from which one reads the relations
\be
T=\left( \frac{\partial \bar E}{\partial \bar s} \right)_{L_x, L_y, L_z, \bar a} \sac
L_x L_y P_z =-\left( \frac{\partial \bar E}{\partial L_z} \right)_{\bar s, L_x, L_y, \bar a} \,,\,\,\,  \ldots \,\,\sac
\bar \Phi=\left( \frac{\partial \bar E}{\partial \bar a} \right)_{\bar s, L_x, L_y, L_z} \,.
\ee
Now we apply the usual extensivity argument to $\bar E=\bar E(\bar s, L_x, L_y, L_z, \bar a)$. For this we first recall Euler's theorem. This states that if a function of multiple variables $f=f(\omega_1, \ldots, \omega_n)$ obeys the scaling law 
$f(k^{\alpha_1} \omega_1, \ldots, k^{\alpha_n} \omega_n) = k^p 
f(\omega_1, \ldots, \omega_n)$, then the function and its partial derivatives are related through
\be
p f(\omega_1, \ldots, \omega_n) = 
\alpha_1 \left( \frac{\partial f}{\partial \omega_1}  \right) \omega_1 + \cdots 
+ \alpha_n \left( \frac{\partial f}{\partial \omega_n}  \right) \omega_n \,.
\ee
In order to apply this to $\bar E$, we note that this obeys the three scaling laws
\bea
\bar E(k \bar s, k L_x, L_y, L_z, \bar a) &=& 
k \bar E(\bar s, L_x, L_y, L_z, \bar a) \,, \nonumber \\
\bar E(k \bar s, L_x, k L_y, L_z, \bar a) &=& 
k \bar E(\bar s, L_x, L_y, L_z, \bar a) \,, \nonumber \\
\bar E(k \bar s, L_x, L_y, k L_z, k \bar a) &=& 
k \bar E(\bar s, L_x, L_y, L_z, \bar a) \,.
\label{EE}
\eea
The first relation can be understood by imagining taking $k$ copies of the original box and `piling them up' along the $x$-direction to make a bigger box. This scales $L_x \ra k L_x$ but leaves $L_y, L_z$ invariant. The total energy and the total entropy of course also get multiplied by $k$. However, $\bar a$ remains invariant because piling up the smaller boxes to make the bigger one along the $x$-direction makes the D7-branes longer, but does not change their number. An analogous argument with $x$ replaced by $y$ leads to the second relation. The third relation can be understood by imagining taking $k$ copies of the original box and `piling them up' now along the $z$-direction. The crucial difference is that in this case the number of D7-branes also gets rescaled as $\bar a \ra k \bar a$. Any other scaling relation obeyed by $\bar E$ can be obtained by composing the three scalings above. 

Applying Euler's theorem, the relations  \eqn{EE} imply 
\bea
\bar E &=& T \bar s - L_x L_y L_z P_x \,, \nonumber \\
\bar E &=& T \bar s - L_x L_y L_z  P_y \,, \nonumber \\
\bar E &=& T \bar s - L_x L_y L_z  P_z  +\bar \Phi \bar a \,.
\eea
It follows that 
\be
P_x = P_y = P_{xy} \sac (P_z - P_{xy})V= \bar \Phi \bar a \,.
\ee
Using \eqn{global} to translate to local densities and the definitions \eqn{F} and \eqn{G} finally leads to the relations \eqn{FF}-\eqn{GG}.


\section{High-temperature analysis}
\label{highApp}

For large values of the temperature, $T\gg a, \mu$, it is possible to find analytic expressions for the metric and the dilaton. To do so, we expand the fields (up to fourth order in $a$) around the black D3-brane solution \eqn{isotropic}:
\bea
\cf(u)&=& 1 - \frac{u^4}{\uh^4} + a^2\hat \cf_2 (u)+ a^4 \hat \cf_4(u) + {O}(a^6) \,,  \\[0.7mm]
\cb(u) &=& 1 + a^2 \hat \cb_2 (u)+ a^4 \hat \cb_4(u) + {O}(a^6) \,, \\[1.7mm]
\phi(u) &=&  a^2 \hat \phi_2 (u)+ a^4 \hat \phi_4(u) + {O}(a^6) \,, 
\label{small_a_exp}
\eea
Note that only even powers can appear because of the symmetry $z\to -z$. We then substitute these expansions into Einstein's equations and solve them order by order in $a$, with the boundary conditions that all the undetermined functions vanish at the boundary $u=0$ (so that we preserve the AdS asymptotics and the condition $\phi_\mt{bdry}=0$), and that, in addition, all the $\hat \cf_n$ vanish at the horizon $u=\uh$. Alternatively, one could simply expand the dilaton as above, solve the third-order dilaton equation \eqn{eq_dil}, and then use eqs.~\eqn{eq_F} and \eqn{eq_B} to obtain $\cf$ and $\cb$. However, the former method allows for an easier implementation of the boundary conditions. 


\subsection{Leading order}

We start by solving for the functions at second order, which turn out to be very simple:
\bea
\hat \cf_2(u)&=& \frac{1}{24 \uh^2}\left[8 u^2( \uh^2-u^2)-10 u^4\log 2 +(3 \uh^4+7u^4)\log\left(1+\frac{u^2}{\uh^2}\right)\right]\,,\cr
\hat \cb_2(u)&=& -\frac{\uh^2}{24}\left[\frac{10 u^2}{\uh^2+u^2} +\log\left(1+\frac{u^2}{\uh^2}\right)\right] \,,\cr
\hat \phi_2(u) &=& -\frac{\uh^2}{4}\log\left(1+\frac{u^2}{\uh^2}\right)\,.
\eea
Evaluating these expressions at the horizon, it is immediate to obtain the temperature of the system to order  $a^2$:
\bea
T= \frac{\cf_1 \sqrt{\cb_\mt{H}}}{4\pi}=\frac{1}{\pi \uh}+\frac{\uh(5\log 2 -2)}{48\pi}a^2+{O}(a^4)\,.
\eea
Inverting this relation we can express the position of the horizon as a function of the temperature:
\bea
\uh = \frac{1}{\pi T}+\frac{5\log 2-2}{48 \pi^3 T^3}a^2 +{O}(a^4)\,.
\label{uha2}
\eea  
This in turn allows us to express the thermodynamical quantities in terms of the physical parameters $a$ and $T$, rather than in terms of $a$ and $\uh$. Using \eqn{uha2} in \eqn{entropy} we see that the entropy density is given by
\bea
s=\frac{\nc^2 e^{-\frac{5}{4}\phi_\mt{H}}}{2\pi \uh^3}=\frac{\pi^2  \nc^2   T^3}{2}+\frac{ \nc^2  T}{16}a^2+{O}(a^4)\,.
\label{entropy_exp}
\eea
The energy and pressures can also be straightforwardly extracted from the asymptotic expansions of the fields. Using the results \eqn{stress_tensor} one obtains
\bea
E&=& -\frac{ \nc^2 }{2\pi^2}\left(\frac{3}{4}{\cal F}_4 +\frac{23}{28}{\cal B}_4\right)+{O}(a^4)=\frac{3\pi^2  \nc^2  T^4}{8}+
\frac{ \nc^2  T^2}{32}a^2+{O}(a^4)\,,\cr
P_{xy}&=& -\frac{ \nc^2 }{2\pi^2}\left(\frac{1}{4}{\cal F}_4 +\frac{5}{28}{\cal B}_4\right)+{O}(a^4)=\frac{\pi^2  \nc^2  T^4}{8}+
\frac{ \nc^2  T^2}{32}a^2+{O}(a^4)\,,\cr
P_z&=& -\frac{ \nc^2 }{2\pi^2}\left(\frac{1}{4}{\cal F}_4 +\frac{13}{28}{\cal B}_4\right)+{O}(a^4)=\frac{\pi^2  \nc^2  T^4}{8}-
\frac{ \nc^2  T^2}{32}a^2+{O}(a^4)\,.
\eea


\subsection{Next-to-leading order}

We now extend the previous analysis to the $O(a^4)$ terms in (\ref{small_a_exp}). This is important because this is precisely the order at which the conformal anomaly \eqn{anomaly} enters.

The full expressions for $\hat \cf_4(u)$, $\hat \cb_4(u)$, and $\hat\phi_4(u)$ are quite cumbersome and not particularly illuminating, so we do not report them here. We limit ourselves to writing down the expressions for the derivatives of $\hat \cb_4(u)$ and $\hat\phi_4(u)$:
\bea
&&\hat \cb_4'(u) =\cr
&&
\hskip .2cm
\frac{u \uh^4}{1152 \left(u^2-\uh^2\right)^2
   \left(u^2+\uh^2\right)^3} \Big[4 u^8 (37 \log 2-19)+4 u^6 \uh^2 (179+64 \log 2)\cr &&
   \hskip 0.4cm -4 u^4 \uh^4 (301+807 \log 2)+3 u^2 \uh^6 (188-952 \log 2)+64
   u^2 \left(u^2+\uh^2\right) \left(u^4-21 \uh^4\right) \log u^2
   \cr &&
   \hskip 0.4cm
    -4 \Big((u^2-\uh^2)^2 \left(u^2+\uh^2\right) \left(u^2+11
   \uh^2\right) \log \uh^2-7 u^2 \uh^2 \left(9 u^4-10 u^2 \uh^2+21 \uh^4\right) \log \left(2 \uh^2\right)
     \cr &&
   \hskip 0.4cm
   -3 \left(54 u^6
   \uh^2+73 u^4 \uh^4+56 u^2 \uh^6+77 \uh^8\right) \log \left(1+u^2/\uh^2\right)
     \cr &&
   \hskip 0.4cm
   +\left(15 u^8+69 u^6 \uh^2-394 u^4
   \uh^4-179 u^2 \uh^6-11 \uh^8\right) \log \left(u^2+\uh^2\right)\Big)\Big]\,,\cr
&&\cr
&& \hat\phi_4'(u)
=
\cr
&& \hskip .2cm
\frac{u \uh^4}{48 (u^2-\uh^2) \left(u^2+\uh^2\right)^2}  \Big[16 u^2 \left(u^2+\uh^2\right) \log u^2
-\left(16 u^4+9 u^2 \uh^2\right) \log \left(u^2+\uh^2\right)  \cr
&&\hskip 0.4cm 
 +\left(u^4 (3+37
   \log 2)-7 u^2 \uh^2 \log \left(2 \uh^2\right)+u^2\uh^2 (34 \log 2-3)\right)
   \cr &&
   \hskip 0.4cm  -\left(18 u^2 \uh^2+21 \uh^4\right) \log
   \left(1+u^2/\uh^2\right)\Big]\,,
\eea
where some constants of integrations have been fixed by requiring regularity at the horizon, and to writing down the equation for 
$\hat{\cal F}_4$, which can be recast in the form
\bea
&& \frac{d\left[ \hat{\cal F}_4(x)/x^4 \right]}{dx} =
\cr 
&& \hskip .2cm
\frac{1}{288 x^5 \left(x^2-1\right) \left(x^2+1\right)^2}
\Big[x^{10} (19-189 \log 2) -x^8 (29+189 \log 2)
   \cr && \hskip .4cm
+x^6(13-211\log 2-48 \log^2 2)+x^4 (9-81\log 2-48 \log^2 2) 
\cr && \hskip .4cm
 -x^2\left(12-30\log 2 -48 \log^2 2\right) +48 \log ^2 2 -32\left(x^2+1\right) \left(7 x^4+3\right) x^4 \log x
     \cr && \hskip .4cm
     + \left(x^2-1\right) \left(x^2+1\right)^2\Big(24 \text{Li}_2 x^4
     -96  \text{Li}_2\frac{x^2+1}{2}+8\pi^2  +96 \log (2 x)\log \left(1-x^4\right)
     \cr && \hskip 4cm
     -96 \log(1-x^2)\log(1+x^2)-9\log^2(1+x^2)\Big) 
     \cr && \hskip .4cm
   + \Big(63 x^{10}+333 x^8+(415+126 \log 2)
   x^6-(75-126 \log 2)x^4
   \cr && \hskip 4cm 
   -(108+126 \log 2)x^2+ (12-126 \log 2)\Big)\log \left(x^2+1\right)\Big]\,,
\eea
with $x \equiv u/\uh$.

We also write down the values taken by all these functions at the horizon
\bea
\hat \cf_{4}(\uh)&=& 0 \,,\cr
&&  \,\cr
\hat \cb_4(\uh)&=& \frac{369-8\pi^2+912 \log 2+354 (\log 2)^2}{6912} \uh^4 \,,\cr
\hat \phi_4(\uh) &=& \frac{3-2\pi^2-6\log 2+87(\log 2)^2}{288}\uh^4\,,
\eea
and their asymptotic expansion near the boundary
\bea
\hat \cf_4(u)&=&  \frac{1}{3456}\left(-915-40\pi^2+1611 \log 2+1440(\log 2)^2+2016 \log\frac{u}{\uh}\right)u^4+{O}(u^6)\,,\cr
\hat \cb_4(u)&=&  \frac{1}{1152}\left(551-567 \log 2-672 \log\frac{u}{\uh}\right)u^4+{O}(u^6)\,,\cr
\hat \phi_4(u)&=& \frac{1}{192}\left(32-27 \log 2-32 \log\frac{u}{\uh}\right)u^4+{O}(u^6)\,.
\eea
Following the same logic as in the previous section, we can compute the temperature at  $O(a^4)$ and solve for $\uh$ obtaining
\bea
\uh = \frac{1}{\pi T}+\frac{5\log 2-2}{48 \pi^3 T^3}a^2 +\frac{180+40\pi^2 -12\log 2-273(\log 2)^2}{13824 \pi^5 T^5}a^4
+{O}(a^6)\,.
\label{uhforT_4order}
\eea
The entropy (\ref{entropy_exp}) receives now an extra contribution
\bea
s=\frac{\pi^2  \nc^2   T^3}{2}+\frac{ \nc^2  T}{16}a^2-\frac{ \nc^2 }{48\pi^2 T}a^4+{O}(a^6)\,,
\label{sss1}
\eea
while the asymptotic coefficients ${\cal F}_4$ and ${\cal B}_4$ defined right below eqn.~\eqn{uOFv} become
\bea
{\cal F}_4 &=& -\pi^4 T^4 -\frac{9 \pi^2 T^2}{16}- \left[ \frac{101}{384}
-\frac{7}{12}\log \left(\frac{2\pi T}{a} \right)
-\frac{7}{12}\log \left(\frac{a}{\mu} \right)\right] a^4 +{O}(a^6)\,,\cr
{\cal B}_4&=& \frac{7\pi^2T^2}{16}a^2+\left[\frac{593}{1152}
-\frac{7}{12}\log \left(\frac{2\pi T}{a} \right)
-\frac{7}{12}\log \left(\frac{a}{\mu} \right)\right]a^4 +{O}(a^6)\,,
\label{F4B4_T}
 \eea
where we have introduced the reference scale $\mu$ required by dimensional analysis and the presence of the anomaly --- see the discussion in Section \ref{stress-section}.

Substituting these expressions in (\ref{stress_tensor}) we find that the energy and pressures are
\bea
E&=& 
\frac{3\pi^2  \nc^2  T^4}{8}+\frac{ \nc^2  T^2 a^2}{32}
+\frac{\nc^2 a^4}{1536 \pi^2} \left[8 \cs -41
+32\log\left(\frac{2\pi T}{a}\right)+32\log\left(\frac{a}{\mu}\right)
\right]+{O}(a^6)\,,\,\,\,\,
\cr
P_{xy}&=&
\frac{\pi^2  \nc^2  T^4}{8}+\frac{ \nc^2  T^2 a^2}{32}-
\frac{\nc^2 a^4}{1536 \pi^2} \left[8 \cs -9
+32\log\left(\frac{2\pi T}{a}\right)+32\log\left(\frac{a}{\mu}\right)
\right]+{O}(a^6)\,,
\cr
P_z&=&
 \frac{\pi^2  \nc^2  T^4}{8}-\frac{ \nc^2  T^2 a^2}{32}+
\frac{\nc^2 a^4}{512 \pi^2} \left[8 \cs -9
+32\log\left(\frac{2\pi T}{a}\right)+32\log\left(\frac{a}{\mu}\right)\right]+{O}(a^6)\,. \,\,\,\,\,\,\,\,\,
\label{en_press_4order}
\eea
Eqs.~\eqn{sss1} and \eqn{en_press_4order} are precisely the results quoted at the beginning of Section \ref{sec-phasediagram}.


\subsection{Check of thermodynamical relations}

First of all, we see that (\ref{en_press_4order}) reproduces the conformal anomaly
\bea
\vev{T_{i}^i}=-E+2P_{xy}+P_z= \frac{ \nc^2 }{48\pi^2}a^4
\eea
regardless of the choice of scheme, i.e.~irrespectively of the values of  $\mu$ and $\cs$.

The free energy $F=E-Ts$ takes the form
\be
F=-\frac{\pi^2  \nc^2  T^4}{8}-\frac{ \nc^2  T^2 a^2}{32}
+\frac{\nc^2 a^4}{1536 \pi^2} \left[8 \cs -9
+32\log\left(\frac{2\pi T}{a}\right)+32\log\left(\frac{a}{\mu}\right)
\right]+{O}(a^6)\,,
\ee
which clearly satisfies \eqn{sphi1} and \eqn{FF}. From the free energy we compute the chemical potential through \eqn{sphi2}:
\bea
\Phi= -\frac{ \nc^2  T^2 a}{16}
+\frac{\nc^2 a^3}{384 \pi^2} \left[8 \cs -41
+32\log\left(\frac{2\pi T}{a}\right)+32\log\left(\frac{a}{\mu}\right)
\right] +{O}(a^5)\,.
\label{PHI_a}
\eea
and with this it is straightforward to verify that \eqn{deltap} holds.

In summary, we see  that all the thermodynamic relations are satisfied regardless of the choice of scheme despite the fact that the free energy, the pressures and the chemical potential are scheme-dependent.


\subsection{Scheme choice}
We conclude this Appendix by showing how the scheme choice adopted in the main text, $\cs=-1$, follows from the requirement that the definitions \eqn{potential} and \eqn{sphi2} agree with one another.

We start from the field strength  $dC(u)$ 
(cf.~eqn.~\eqn{fieldstrength}) 
\be
\partial_u C_{txy} (u)= a\frac{e^{\frac{7}{4}\phi}\sqrt{\cb}}{u^3}\,.
\ee
Using the results above this equation can be integrated to next-to-leading order in $a$. We fix the integration constant so that  $C_{txy}(\uh)=0$ and use  (\ref{uhforT_4order}) to express the result in terms of $T$ and $a$. Expanding the result near the boundary and transforming to the FG coordinate $v$ through (\ref{uOFv}), we obtain the asymptotic fall-off \eqn{ctxy} with
\be
\cc=
\frac{1}{48}\left( 24a\pi^2T^2+17 a^3 \right) 
-\frac{2}{3} a^3 \log \left( \frac{2\pi T}{a} \right) 
-\frac{2}{3} a^3 \log \left( \frac{a}{\mu} \right)  + O(a^5) \,.
\ee
Requiring that substitution of this expression in eqn.~\eqn{potential} result exactly in (\ref{PHI_a}) fixes both the normalization chosen in \eqn{potential} and $\cs=-1$.


\section{Low-temperature analysis}
\label{lowApp}

In this appendix we consider the limit $T\ll a$, which is also amenable to analytical treatment. The key observation is that at $T=0$ we must have \mbox{$-g_{tt}=g_{xx}=g_{yy}$} and therefore 
$\cf\cb=1$, since in this limit the Lorentz symmetry in the $t,x,y$ directions must be recovered. Either from the expansion of the metric coefficients (\ref{metricExpansion}) or from the functions themselves \eqn{expansionFunctions}, we see that this implies a relation between the undetermined coefficients:
\bea
{\cal B}_4(a,0) +{\cal F}_4(a,0)= \frac{121}{576}a^4\,.
\eea
Using this to eliminate ${\cal B}_4(a,0)$, eqs.~\eqn{stress_tensor} reduce in this limit to
\bea
E(a,0)&=&\frac{28\cs-1}{5376 \pi^2} \nc^2  a^4+\frac{ \nc^2 }{28\pi^2}{\cal F}_4(a,0)\,,\cr
P_{xy}(a,0)&=&-\frac{28\cs-1}{5376 \pi^2} \nc^2  a^4-\frac{ \nc^2 }{28\pi^2}{\cal F}_4(a,0)\,,\cr
P_z(a,0)&=&\frac{84\cs+109}{5376 \pi^2} \nc^2  a^4+\frac{3  \nc^2 }{28\pi^2}{\cal F}_4(a,0)\,,
\label{en_press_low_exp}
\eea
from which it follows immediately that
\bea
E(a,0)=-P_{xy}(a,0) \sac 
P_z(a,0)=3E(a,0)+\frac{ \nc^2 }{48 \pi^2}a^4\,.
\eea
Looking at the region near the origin in Fig.~\ref{enpress_aoverT} one can verify that these relations are satisfied by our numerical solution. Note also that the first relation is consistent with the fact that $s=0$ in this limit, since it implies that $F=E$. Thus we can compute the chemical potential as usual by differentiating with respect to $a$ either of these quantities:
\bea
\Phi(a,0)=\left(\frac{\partial F(a,0)}{\partial a}\right)_T = 
\frac{28\cs-1}{1344 \pi^2} \nc^2  a^3+\frac{ \nc^2 }{28\pi^2}\frac{\partial {\cal F}_4(a,0)}{\partial a}\,.
\eea
Requiring that this $\Phi(a,0)$ also satisfies the relation $P_z(a,0)-P_{xy}(a,0)=\Phi(a,0) a$ implies the differential equation
\bea
a \frac{\partial {\cal F}_4(a,0)}{\partial a}-4 {\cal F}_4(a,0)-\frac{7}{12}a^4=0\,,
\eea
which can be easily solved up to an integration constant
\bea
{\cal F}_4(a,0)=\frac{7}{12}a^4 \log \left( \frac{a}{\mu} \right) 
+c_\mt{int} a^4\,,
\eea
where as usual we have introduced the necessary scale $\mu$. Substituting this into (\ref{en_press_low_exp}) one recovers the $T$-independent parts of (\ref{E_low})-(\ref{pzlow}).

So far we have strictly set $T=0$, but we can also compute analytically the first correction in $T/a$. We know in fact that when 
$a\gg T$ our solution reproduces the Lifshitz geometry found in \cite{ALT}, and therefore that the entropy density must scale as (see the discussion below \eqn{sss})
\bea
s(a,T\ll a)=c_\mt{ent} \nc^2  a^{1/3} T^{8/3}+\cdots\,,
\label{s_low}
\eea 
where $c_\mt{ent}\simeq 3.2$ is a constant that can be read off from the plot of the entropy in Fig.~\ref{scalings} and the ellipsis refers to terms of higher order in $T/a$. Integrating this expression with respect to $T$ yields the free energy
\bea
F(a,T\ll a) = -\frac{3 c_\mt{ent}}{11} a^{1/3} T^{8/3}+\cdots\,.
\eea
From this and (\ref{s_low}) one can easily obtain all other thermodynamical quantities at leading order in $T/a$. This completes the derivation of (\ref{E_low})-(\ref{pzlow}).


\end{document}